
\overfullrule=0pt
\magnification=1100
\baselineskip=4ex
\raggedbottom
\font\eightpoint=cmr8
\font\sevenpoint=cmr7
\font\fivepoint=cmr5
\headline={\hfill{\fivepoint  EHLML-21/Sept/92}}
\def\R{{\rm I\kern -1.6pt{\rm R}}}
\def\Z{{\bf Z}}
\def\C{{\rm I\kern -6.0pt{\rm C}}}
\def\mod{{\rm mod}}
\def\lanbox{{$\, \vrule height 0.25cm width 0.25cm depth 0.01cm \,$}}
\def\L{{\cal L}}
\def\D{{\cal D}}
\def\F{{\cal F}}
\def\spec{{\rm spec}}
\def\d{{\rm d}}
\def\Pf{{\rm Pf}}
\def\Tr{{\rm Tr}}
\def\mfr#1/#2{\hbox{${{#1} \over {#2}}$}}
\def\const.{{\rm const.}}
\catcode`@=11
\def\eqalignii#1{\,\vcenter{\openup1\jot \m@th
\ialign{\strut\hfil$\displaystyle{##}$&
        $\displaystyle{{}##}$\hfil&
        $\displaystyle{{}##}$\hfil\crcr#1\crcr}}\,}
\catcode`@=12
\centerline{\bf FLUXES, LAPLACIANS AND KASTELEYN'S THEOREM}
\bigskip
\bigskip
\halign{\qquad#\hfil\qquad\qquad\hfil&#\hfil\cr
Elliott H. Lieb \footnote{$^*$}{\baselineskip=12pt \sevenpoint
Work partially supported by U.S. National Science Foundation grant
PHY90-19433A01}
&Michael Loss \footnote{$^\dagger$}{\baselineskip=12pt
\sevenpoint Work partially supported by U.S. National Science Foundation
grant DMS92-07703} \cr
Departments of Mathematics and Physics
&School of Mathematics \cr
Princeton University
&Georgia Institute of Technology \cr
P.O. Box 708, Princeton, NJ  08544-0708
&Atlanta, GA  30332-0160 \cr }
\bigskip
\bigskip
\vskip 0.5truein
{\bf Abstract:\/} The following problem, which stems from the ``flux
phase'' problem in condensed matter physics, is analyzed and extended
here: One is given a planar graph (or lattice) with prescribed
vertices, edges and a weight $\vert t_{xy}\vert$ on each edge $(x,y)$.
The flux phase problem (which we partially solve) is to find the real
phase function on the edges, $\theta(x,y)$, so that the matrix
$T:=\{\vert t_{xy}\vert {\rm exp}[i\theta(x,y)]\}$ minimizes the sum of
the negative eigenvalues of $-T$. One extension of this problem which
is also partially solved is the analogous question for the
Falicov-Kimball model.  There one replaces the matrix $-T$ by $-T+V$,
where $V$ is a diagonal matrix representing a potential.  Another
extension of this problem, which we solve completely for planar,
bipartite graphs, is to maximize $\vert {\rm det}\ T \vert$.  Our analysis
of this determinant problem is closely connected with Kasteleyn's 1961 theorem
(for arbitrary planar graphs) and, indeed, yields an alternate, and we
believe more transparent proof of it.
\vfill\eject
\bigskip\noindent
{\bf I.  INTRODUCTION}

The genesis of this paper was an attempt to understand a problem in
condensed matter physics related to questions about electron correlations,
superconductivity and electron-magnetic field interactions.  The basic
idea, which was proposed a few years ago is that a magnetic field
can lower the energy of electrons when the electron density is not small.
Certain very specific and very interesting mathematical conjectures about
eigenvalues of the Laplacian were made and the present paper contains a
proof of some of them.  Furthermore, those conjectures lead to additional
natural conjectures about determinants of Laplacians which we both present
and prove here.  It is not clear whether these determinantal theorems have
physical applications but they might, conceivably in the context
of quantum field theory.  Some, but not all, of the results given here were
announced earlier in [LE].

The setting is quantum mechanics on a graph or lattice.  (All our
terminology will be defined precisely in the sequel.)  Physically, the
vertices of our graph $\Lambda$ can be thought of either as a
discretization of space (i.e. replace the Laplacian by a finite difference
operator) or they can be seen as locations of atoms in a solid.  There are
$\vert \Lambda \vert$ vertices.  In the
atomic interpretation the edges become electron bonds joining the atoms,
and the model is known as the tight-binding model or H\"uckel model.  The
natural Laplacian $\L$ associated with $\Lambda$ is a $\vert \Lambda \vert
\times \vert \Lambda \vert$ matrix indexed by the
vertices of $\Lambda$ and whose diagonal elements satisfy $-\L_{xx} =$
number of attached edges (or valency) of vertex $x$.  The other
elements are $\L_{xy} = 1$ if $x$ and $y$ are connected by an edge, and
zero otherwise.

For us it is more convenient to consider the matrix $\widehat \L$ which is
the Laplacian without the diagonal term, i.e., $\L_{xx}$ is replaced by
zero.  In the context of graph theory $\widehat \L$ is also known as the
adjacency matrix.  There are three excuses for this:
(i) in the solid state context
$\widehat \L$ is the natural object because atoms do not bond to
themselves; (ii) most of the graphs that are considered in the physics
literature have constant valency, so $\widehat \L$ and $\L$ have
the same spectrum modulo a constant which is equal to this valency; (iii)
mathematically $\widehat \L$ seems to be the more natural object --- from
our point of view, at least --- because its spectrum on a bipartite graph
is always a union of pairs $\lambda$ and $-\lambda$ (when $\lambda \not=
0$), as explained in Sect.~II.  The spectrum of $\L$ generally does not have
any such symmetry.

We label the eigenvalues of $\widehat \L$ by $\lambda_1 (\widehat \L) \geq
\lambda_2 (\widehat \L) \geq \dots$.  The Hamiltonian for a single electron
is $-\L$ or $-\widehat \L$, and we take it to be $-\widehat \L$ here.  If
our system has $M$ {\it free\/} electrons the rule of quantum mechanics is
that the eigenvalues of our system are all the possible sums of $M$ of the
$(-\lambda_i$)'s in which each $-\lambda_i$ is allowed to appear at most
twice in the sum.  There are ${2 \vert \Lambda \vert \choose M}$
eigenvalues.  In particular, if $M = 2N$ the smallest eigenvalue is
$$E^{(N)}_0 = - 2 \sum \limits^N_{j=1} \lambda_j (\widehat \L). \eqno(1.1)$$

A (spatially varying) magnetic field is now added to the system in the
following way.  $\widehat \L_{xy}$ is replaced by $T_{xy} =
\widehat \L_{xy} \exp [i
\theta (x,y)]$, with $\theta$ real and with $\theta (x,y) = - \theta (y,x)$
so that $T$ is Hermitian.  The function $\theta (x,y)$ is interpreted
physically
as the integral of a magnetic vector potential from the point $x$ to the
point $y$.  This $T$ is the discrete analogue of replacing the Laplacian on
$\R^n$ by $(\nabla - i A(x))^2$ (with $\nabla =$ gradient), which is the
Laplacian on a $U(1)$ bundle.

The central question that we address is this:  {\it What choice of $\theta$
minimizes $E^{(N)}_0$ for a given $N$?}

In order to appreciate this question, consider the $N = 1$ case.  Then,
$\theta \equiv 0$ is an answer because (with $\phi$ being the normalized
largest eigenvector of $T$) $\lambda_1 (T) = \sum \overline \phi_x \phi_y
\widehat \L_{xy} \exp [i \theta (x,y)]$
{$\leq \sum \vert \phi_x \vert \vert
\phi_y \vert \widehat \L_{xy} \leq \lambda_1 (\widehat \L)$}.  This proof
that $\theta \equiv 0$ is optimum also works in a more general setting,
namely for the lowest eigenvalue of the ``Schr\"odinger operator'' $-T +
V$, where $V$ is any real diagonal matrix.  Again, $T = \widehat \L$, or
$\theta \equiv 0$, minimizes $-\lambda_1 (T - V)$.  The same is true
in $\R^n$ for $- (\nabla - i A(x))^2 + V(x)$; the minimum occurs when $A(x)
\equiv 0$.  This conclusion is known as the {\it diamagnetic inequality\/}
and states, physically, that ``a magnetic field raises the energy''.

It was
discovered by [AM] and [KG] that the situation can be quite different when
$N$ is close to $\mfr1/2 \vert \Lambda \vert$.
(When $N = \vert \Lambda \vert, E^{(N)}_0 = \Tr T = 0$
for all $\theta$; hence $N = \mfr1/2 \vert \Lambda \vert$ is the most
extreme case.)  Since then, the problem has been investigated for
various lattices and $N$'s by several authors such as [BR], [BBR], [HLRW],
[RD], [WP], [WWZ] some of whom consider it to be
important in the theory of high
temperature superconductivity.  [HLRW], for example, start with the square
lattice $\Z^2$, take $\Lambda$ to be a large rectangular subset of $\Z^2$,
and then let $\vert \Lambda \vert \rightarrow \infty$ and $N \rightarrow
\infty$ with $N/\vert \Lambda \vert$ fixed.  They also take the magnetic
flux (which is the sum of the $\theta$'s around the edges of a face, and
which is defined in Sect.~II) to have the {\it same value\/} in each square
box of $\Z^2$.  On the basis of their numerical evidence they proposed that
flux/box $= 2 \pi N /\vert \Lambda \vert$ is the optimal choice.
In [AM] the term ``flux phase'' was introduced to describe this state in
which the presence of a magnetic field lowers the energy.

It should be pointed out that the spectrum of $\widehat \L$ for $\Z^2$ as a
function of constant flux/box was discussed by many authors for
many years; it was
Hofstadter [HD] who grasped the full beauty of this object --- which is
anything but a continuous function of the flux and which is full of gaps
--- and called it a ``butterfly''.  The spectrum can be found by solving a
one-dimensional difference equation, due to Harper [HP], which is a
discrete analogue of , but more complicated than, Mathieu's equation.  The
spectrum is such a complicated function of the flux that it is difficult to
decide on the optimum flux for a given $N$.

The most striking case is $N/\vert \Lambda \vert = \mfr1/2$, or $M = \vert
\Lambda \vert$, which is
called the {\bf half-filled band}.  The optimal flux is supposed to be
$\pi$, which is the maximum possible flux since flux is determined only
modulo $2 \pi$ and since flux and $-$flux yield identical spectra.  It is this
case that we investigate in this paper in an attempt to verify the rule
just stated and which appears in
[AM], [HLRW], [RD].  We are completely successful only in some special
cases, but we have been able to generalize the problem in several
interesting directions.  For example, one of our main results is Theorem
3.1.  It completely solves the problem for determinants (i.e. for products
of eigenvalues instead of sums of eigenvalues) on bipartite planar graphs.

Our determinant theorem turns out to be closely related to Kasteleyn's
famous 1961 theorem about planar graphs, which allowed him to solve
(in principle) the
dimer problem and Ising model for all planar graphs.  Our
route, via fluxes, gives an alternative proof of Kasteleyn's theorem and,
we believe, a more transparent one.  This is presented in the Appendix.

The setting we adopt is a general graph $\Lambda$, with no particular
symmetry such as $\Z^2$ enjoys, and an {\it arbitrary}, but {\it fixed\/}
amplitude
$\vert t_{xy} \vert > 0$ given on each edge ($\vert t_{xy} \vert = 1$ in
the case of
$\widehat \L$\ ).  The problem is to determine $\theta$ and
$T:= \{ t_{xy} \}_{x,y \in
\Lambda}$ with $t_{xy} = \vert t_{xy} \vert \exp [ i \theta (x,y)]$ so as
to minimize the (absolute) {\bf ground state energy}
$$E_0 (T) := - \Tr \vert T \vert - \Tr T = - \Tr \vert T \vert,
\eqno(1.2)$$
with $\Tr =$ Trace.  The right side of (1.2) is twice the sum of the
negative eigenvalues of $-T$.  For a bipartite graph this is the sum of the
$\vert \Lambda \vert /2$ or $(\vert \Lambda \vert - 1)$/2 lowest
eigenvalues of $-T$.

A word has to be said here about different definitions of ground state
energy.  Electrons have two spin states available to them and the Pauli
exclusion principle states that each eigenstate of $-T$ can be occupied by
at most one electron of each kind.  Thus, each eigenstate can be occupied
by 0 or 1
(twice) or 2 electrons.  That explains the factor of 2 in (1.1):  there the
lowest $N$ eigenstates of $-T$ are each occupied by two electrons.
Our definition
of $E_0 (T)$ in (1.2) is the absolutely lowest ground state energy and
corresponds to the electron number being twice the number of negative
eigenvalues.  On the other hand, the half-filled band would have the
electron number equal to $\vert \Lambda \vert$ by definition.  If $\vert
\Lambda \vert = 2N$ is even, then the half-filled band ground state energy
is given by (1.1) with $N = \vert \Lambda \vert /2$.
If $\vert \Lambda \vert = 2N + 1$, the half-filled band
ground state energy is $-2 \sum\nolimits^N_{j=1} \lambda_j -
\lambda_{N+1}$.  It is this half-filled band energy that is mostly
considered in the
physics literature.  However, we regard our definition (1.2) as
mathematically more natural and physically as interesting as the strict
half-filled band definition.  {\it For bipartite graphs $E_0 (T)$ and
$E^{(N)}_0$ with $N = \vert \Lambda \vert /2$ or $(\vert \Lambda \vert -
1)/2$ agree with each other.}  (Note that if $\Lambda$ is bipartite and
$\vert \Lambda \vert = 2N + 1$ then $\lambda_{N + 1} = 0$.)

The two definitions can produce strikingly different conclusions, however,
in special cases.  In [RD] the ground state energy $E_0 (T)$ of $N$
electrons (including spin) hopping on a ring of $N$ sites is considered.
By Theorem 4.1 we know that for a ring with $N$ odd (and which is therefore
not bipartite), the expression (1.2)
is minimized by flux $\pi$ and flux 0.  However it has been shown in some cases
(see [RD]) that the half-filled band energy for such a ring is minimized by
the flux $\pi /2$ (which, incidentally, we call the canonical flux in
this paper).

There is an important difference between our minimization problem and the
one in [HLRW] and some other papers in the physics literature.  For a
regular structure like $\Z^2$ we allow {\it different fluxes in different
boxes}.  In the physics literature the problem is sometimes stated with
{\it constant fluxes\/} or with periodic fluxes.  We find our formulation
(with arbitrary fluxes) to be
more natural mathematically and we believe it to be more natural in those
physical problems where this theory might be applicable.

Besides the ground state energy problem we consider other functions of $T$,
such as $\ln \vert \det T \vert = \Tr \ln \vert T \vert$.  A particularly
important one, physically,
is $\ln \Xi$ where $\Xi$ is the {\bf grand canonical partition
function} with chemical potential $\mu$ and inverse temperature $\beta$,
given by
$$\Xi = \sum \limits_{m_1} \cdots \sum \limits_{m_{\vert \Lambda \vert}}
\sum \limits_{n_1} \cdots \sum \limits_{n_{\vert \Lambda
\vert}} \exp \left[ \beta \sum \limits^{\vert \Lambda \vert}_{j=1} \lambda_j
(n_j + m_j) + \beta \mu \right]
= \prod \limits^{\vert \Lambda \vert}_{j=1} \big\{ 1 + \exp [\beta
(\lambda_j + \mu)] \big\}^2, \eqno(1.3)$$
where the sum on each $n_i$ and $m_i$ is over the set $\{ 0,1\}$.
The physical {\bf free energy} is defined by $\F = - \beta^{-1} \ln
\Xi$.  We consider only $\mu = 0$ here because that corresponds to a
half-filled band in the bipartite case (see (4.5) and footnote).

Another important quantity is the {\bf gap}, $G(T)$, which is {\it not\/}
defined by a trace.  We define it to be
$$G(T) = - \lambda_{N+1} + \lambda_N ,\eqno(1.4)$$
where $N$ is the number of negative eigenalues of $-T$, i.e., the smallest
number such that $E^{(N)}_0 = E_0 (T)$.  Clearly $G(T)$
is the energy needed to add one more electron to the system from the
absolute ground state.  For the
half-filled band on a bipartite graph with $\vert \Lambda \vert = 2N$,
$G(T) = 2 \lambda_N$.  This, however, may not be mathematically interesting
because $\lambda_N$ may be automatically zero for dimensional reasons.
That is, if $\vert A \vert$ and $\vert B \vert$ are the two subsets of
vertices of $\Lambda$ that define the bipartite structure, then $T$ always
has at least $\big\vert \vert B \vert - \vert A \vert
\big\vert$ zero eigenvalues.
For this reason we define $\widetilde G(T)$ for a {\it bipartite\/} $\Lambda$
(with $\vert \Lambda \vert$ odd or even)
to be
$$\widetilde G(T) = \lambda_{\vert A \vert} - \lambda_{\vert B \vert +
1}, \eqno(1.5)$$
assuming $\vert B \vert \geq \vert A \vert$.  We can then ask the question:
{\it Which flux maximizes $G(T)$ or $\widetilde G(T)$ in the bipartite
case?}

So far we have discussed free --- or noninteracting --- electrons.  The
same questions can be asked for interacting electrons and very much less is
known in that case.  In Sect.~VIII, however, we are able to carry over our
techniques to one example --- the Falicov-Kimball model.

Many of these results were announced in [LE].  We thank P. Wiegmann for
bringing this problem to our attention and along with I. Affleck, D.
Arovas, J. Bellissard and J. Conway, for helpful discussions.
\bigskip\noindent
{\bf II.  DEFINITIONS AND PROPERTIES OF FLUXES}

A {\bf graph} $\Lambda$ is a finite set of {\bf vertices} (or {\bf sites}),
usually denoted by lower case roman letters $x,y,z$ etc., together with {\bf
edges} (or {\bf bonds}), which are certain unordered pairs of {\it
distinct\/} sites and are denoted by $(x,y)$, equivalently $(y,x)$.  Thus
there will be at most one edge between two vertices.  The set of sites or
vertices will be denoted by $V$ or $V(\Lambda)$ and the number of them by
$\vert \Lambda
\vert$.  The set of edges will be denoted by $E$ or $E(\Lambda)$. If
$(x,y) \in E$ the
sites $x$ and $y$ are said to be {\bf end points} of the edge $(x,y)$.

A graph $\Lambda$ is connected if for every pair of sites $x$ and $y$ there
is a {\bf path} $P$ in $\Lambda$ connecting $x$ and $y$, i.e., there is a
sequence of points $x = x_0, x_1, x_2, \dots , x_n = y$ such that $(x_i,
x_{i+1})$ is an edge for every $0 \leq i < n$. Although $\Lambda$ is
not just the set of vertices, but also contains the edges, we
shall nevertheless sometimes write $x \in \Lambda$ where $x$ is a
site in $\Lambda$.

A {\bf hopping matrix} $T$ associated with a graph $\Lambda$ is a Hermitian
$\vert \Lambda \vert \times \vert \Lambda \vert$ matrix indexed by the
sites of $\Lambda$, with elements denoted by $t_{xy} = \overline{t_{yx}}$
for $x,y \in \Lambda$, and with the important property that $t_{xy} \not= 0$
only if $(x,y) \in E$, i.e., if $x$ and $y$ are connected by an edge.  In
particular, $t_{xx} = 0$ for all $x \in V$.  The $T$ matrix is the important
object here.  For that reason if $t_{xy} = 0$
for any edge $(x,y)$ we might as well delete this edge from the graph
$\Lambda$.  Thus, without loss of generality we can assume that every
$t_{xy}$ is nonzero and that the corresponding graph $\Lambda$ is
connected.  If it is not connected $T$ breaks up into blocks which can be
considered separately.  We call $\vert t_{xy} \vert$ the {\bf hopping
amplitudes}.  No other assumption is made about $t_{xy}$ unless explicitly
stated otherwise.  The eigenvalues of $T$ are usually denoted by $\lambda$,
and sometimes by $\lambda (T)$ to be more specific.

A {\bf circuit} $C$ of length $\ell$ in $\Lambda$ is an ordered sequence of
{\it
distinct\/} sites $x_1, \dots, x_\ell$ with the property that $(x_i, x_{i+1})$
is an edge for $i = 1, \dots , \ell$ with $x_{\ell+1} \equiv x_1$.  We
explicitly {\it include\/} $\ell = 2$.
Note that $x_2, \dots, x_\ell, x_1$ is the same circuit as $C$, but
$x_\ell, x_{\ell-1}, \dots, x_1$ is different.

If $C$ is a circuit then we can define the {\bf flux} $\Phi_C$ {\bf of} $T$
{\bf through} $C$, which is a number in $[0,2\pi)$, as

$$\Phi_C \equiv \arg \left( \prod \limits^n_{i=1} t_{x_i, x_{i+1}} \right)
\equiv \arg \left( \prod \limits_C T \right). \eqno(2.1)$$
The symbol $\prod \limits_C T$ has an evident meaning.

A {\bf gauge transformation} is a diagonal unitary transformation $U$ with
elements, $u_{xy} =
\exp [i \phi_x] \delta_{xy}$, where $\phi_x : V \rightarrow \R$ is a
function on the sites.
Obviously a gauge transformation $T \rightarrow U^* T U$ leaves the
spectrum of $T$ and all the
fluxes unchanged.

{\bf 2.1.  LEMMA (Fluxes determine the spectrum).}  {\it Let $T$ and
$T^\prime$ be two hopping matrices which have the same hopping
amplitudes and the same flux through each circuit $C$ of the graph
$\Lambda$.  Then there is a gauge transformation $U$ such that
$T^\prime = U^*TU$.}

{\it Proof:}  By our convention the $t^{\phantom{\prime}}_{xy}$ and
$t^\prime_{xy}$ are never zero.  Thus $W_{xy} \equiv
t^{\phantom{\prime}}_{xy}/t^\prime_{xy}$ satisfies $\vert W_{xy} \vert = 1$
for all edges $(x,y)$, and the flux of $W$ through each circuit of $\Lambda$ is
zero.  Let $x_0$ be an arbitrary but henceforth fixed site in $\Lambda$.  For
any $x$ we can pick a path $P$ connecting $x_0$ and $x$ and define $\phi_x
= \arg \left( \prod \limits_P W \right)$.  The value of $\phi_x$ does not
depend on the choice of the path because, if $P^\prime$ is another path
connecting $x_0$ and $x$, we have that $\arg \left( \prod \limits_P W
\right) = \arg \left( \prod \limits_{P^\prime} W \right)$ since the flux
through the circuit given by connecting $x_0$ to $x$ along $P$ and then
connecting $x$ to $x_0$ along $P^\prime$ is zero.  If $x$ and $y$ are
arbitrary sites in $\Lambda$ with $(x,y) \in E$, and if we take $P_x$
to be a path connecting $x_0$
to $x$ and $P_y$ a path connecting $x_0$ to $y$ we observe that $1 =
\prod \limits_{P_x} W \ W_{xy} \prod \limits_{P_y} \overline{W}$
since $P_x$ followed by $(x,y)$ is a path from $x_0$ to $y$.  But this
equals $\exp [i (\phi_x -
\phi_y)] W_{xy}$ and hence $W_{xy} = \exp [-i (\phi_x - \phi_y)]$.  Thus
$t^\prime_{xy} = \exp [i\phi_x] t_{xy} \exp [-i \phi_y]$, which proves the
lemma.  \lanbox

Up to now a graph has been regarded as an abstract object consisting of
vertices and edges.  Now we wish to regard graphs as embedded either in $\R^2$
or $\R^3$.  This means that the sites of $\Lambda$ can be regarded as
distinct fixed points in $\R^3$ and each edge $(x,y)$ will be identified
with exactly one piecewise linear curve between $x$ and $y \in
\R^3$.  It is convenient to {\it exclude\/} the end points $x$ and $y$ in the
definition of an edge.  We require that any one edge does not intersect the
other edges or the sites. Circuits are then identified with simple, oriented
closed curves.

Obviously any graph can be embedded in $\R^3$ but only some graphs, called
{\bf planar graphs}, can be embedded in $\R^2$.  It is these graphs that
will mostly concern us in this paper.

The set of edges $E$ and sites $V$ of a planar graph, regarded as a set
in $\R^2$, is closed.  Its complement is therefore open and this
complement has a finite number of connected components which we label
$F_0, F_1, F_2, \dots$.  We define $F_0$ to be the open set that
contains the point at infinity, (i.e. the exterior of the graph).  The
others we call the {\bf faces} of the graph $\Lambda$.  Each face has a
boundary and this boundary is composed of a subset of the edges and
sites of $\Lambda$.  For later purposes we call {\bf elementary circuits}
those circuits which are entirely contained in the boundary of a single
face.

If the graph is planar a circuit, $C$, of length greater than 2 will have
an inside and an outside.  The interior, which is an open set, is then
the union of a certain number of faces, edges and vertices called {\bf
interior faces, interior edges} and {\bf interior vertices}.  We denote
their numbers by $f, e, v$. We can speak of  the {\bf orientation} of
$C$ as being either positive (anticlockwise) or negative (clockwise)
according as the winding number with respect to a point in its interior
is either $+1$ or $-1$.

In general, an arbitrary specification of fluxes through the circuits of
$\Lambda$ may be inconsistent in the sense that there may not exist a
choice of $T$ with the prescribed fluxes.  Some kind of divergence --
or closedness condition is needed.  In two dimension, however, the
following lemma shows how fluxes can be specified in a consistent way.

{\bf 2.2.  LEMMA (Construction of phases from fluxes).}  {\it Let
$\Lambda$ be a planar graph and let $F_1, F_2, \dots F_f$ be its
faces.  Let $\Phi^1, \Phi^2, \dots , \Phi^f$ be any given numbers in
$[0, 2 \pi)$.  (We call $\Phi^j$ the {\bf flux through} $F_j$.)  Then
there is a function $\theta (x,y) : E(\Lambda) \rightarrow [0, 2 \pi)$
so that the fluxes defined by (2.1) with $t_{xy} := \exp [i \theta (x,y)]$
satisfy $\Phi_C = \sum\nolimits_{\scriptstyle{interior \ faces \ of \ C}}
\Phi^j$ for every positively oriented
circuit $C$ on $\Lambda$.}

{\it Proof:}  Pick a point $z_j = (z^1_j, z^2_j) \in \R^2$
in each face $F_j$ and consider the vector (really, a one-form)
field on $\R^2$, $\vec A (x) = \vec A (x^1,
x^2)$, with singularities at the $z_j$'s given by
$$\vec A (x) = \sum \limits_{{\rm all \ faces \ } F_j \ {\rm of} \ \Lambda}
\Phi^j \vec a (x-z_j)$$
with $\vec a (x) = (2 \pi)^{-1} [(x^1)^2 + (x^2)^2]^{-1} (-x^2, x^1)$.
For each pair $x,y \in V(\Lambda)$ with $(x,y)
\in E(\Lambda)$ define $\theta (x,y) = \int \limits^y_x
\vec A$, i.e., the integral from $x$ to $y$
along the curve representing the edge $(x,y)$.  The flux $\Phi_C$ through
any circuit $C$ on $\Lambda$ is given by $\oint_C A$.  By Cauchy's integral
formula (or Stokes' theorem), this integral equals $\Phi_C$ given above.
\lanbox

{\it Remark:}  A more practical and direct way to construct phases $\theta
(x,y)$ satisfying the conclusion of Lemma 2.2 is to concentrate the vector
field $\vec a (x - z_j)$ along a line.  More precisely, let $L_j$
denote some semi-infinite line starting from $z_j$ and extending to
infinity, but which does not intersect any of the sites of $\Lambda$ and
whose intersection with an edge is always transverse, i.e., nontangent.  We
orient $L_j$ from $z_j$ outwards.  For each {\it ordered\/} pair of
sites $(x,y)$, with $(x,y) \in E$, first orient the curve representing
$(x,y)$ in the direction from $x \in \R^2$ to $y \in \R^2$.  Call this
oriented curve $\varepsilon$.  Then let $\mu_j (x,y) \equiv
\Sigma_{L_j \cap \varepsilon} (\pm 1)$, where the sum is over all the
points of intersection of $L_j$ with this curve $\varepsilon$ and
where the $+$ sign (resp. $-$ sign) is taken if the (counterclockwise)
angle from $L_j$
to $\varepsilon$ is less than (resp. more than) $\pi$.  Finally, we set
$\theta (x,y) = \Sigma^f_{j=1} \Phi^j \mu_j (x,y)$.

It is a fact that every planar graph, $\Lambda$, can be {\bf triangulated},
i.e., that there is a planar graph $\Lambda^\prime$ with precisely the {\it
same\/}
vertices as $\Lambda$ and whose edges $E^\prime$ (which are now
sets of curves) contain $E$, the edge-set of $\Lambda$, and with the property
that $\Lambda^\prime$ is {\bf triangular}.The concept of a planar graph
$\Lambda^\prime$ being triangular means that every one of the faces of
$\Lambda^\prime$ has as its boundary the union of three edges and
three vertices.  We say that $\Lambda^\prime$ is a {\bf triangulation}
of $\Lambda$.
It is easy to check that the three edges must always form a circuit.  Note
that, in general, a graph $\Lambda$ can be triangulated in several ways.

{\bf 2.3.  LEMMA (Number of triangles in a circuit).}  {\it Let $\Lambda$ be a
triangular planar graph and let $C$ be a circuit in $\Lambda$ of length
$\ell \geq 2$.  Let $f$ and $v$ denote the number of interior faces and
interior vertices of $C$.  Then
$$\ell - f + 2v = 2. \eqno(2.2)$$}

For $\ell = 2$, (2.2) is clearly true (with $f = v = 0$).  Therefore we
need consider only $\ell > 2$.

{\it First proof:}  Let $\beta$ be one of the $\ell$ edges in $C$.  This
edge must be part of the boundary of exactly one of the inner triangles,
which we call $\tau$.  The boundary of $\tau$ contains 3 edges, $\beta_1,
\beta_2, \beta_3$.  There are two cases.
\item{(a)}  $\beta_1 = \beta$ and $\beta_2, \beta_3$ are interior to $C$.
\item{(b)}  $\beta_1 = \beta$, $\beta_2$ is an edge of $C$, $\beta_3$ is
interior to $C$.
\smallskip\noindent
In case (a) we consider the circuit $C^\prime$ whose edges are the same as
those of $C$ except that $\beta$ is replaced by the two edges $\beta_2,
\beta_3$.  In case (b) we remove $\beta_1$ and $\beta_2$ from $C$ and
replace them with $\beta_3$.  It is easy to check that $\ell^\prime -
f^\prime + 2v^\prime = \ell - f + 2v$ and that $f^\prime = f - 1$.  By
successively removing triangles in this way we eventually have only one
triangle left, in which case $\ell = 3, f = 1, v = 0$.

{\it Second proof:}  Euler's formula says that (total number of vertices) $+$
(total number of faces) $-$ (total number of edges) $=1$.  Since $\ell$
also equals
the number of vertices in $C$, we have that $1 = (v + \ell) + (f) - (\ell +
e)$, where $e$ is the number of interior edges.  Each edge in $C$ lies in
the boundary of precisely one interior triangle, while each interior edge
lies in the boundary of two such triangles.  Since each triangle has three
edges in its boundary, and since $C$ has $\ell$ edges, we have $3 f = 2 e +
\ell$.  Therefore $e = 3 f/2 - \ell/2$ and $1 = v + f - e = v - f/2 + \ell
/2$. \lanbox

{\bf 2.4.  COROLLARY ($f$ is independent of triangulation).}  {\it Let
$\Lambda$ be an arbitrary planar graph and let $\Lambda^\prime$ denote a
triangulation of $\Lambda$.  For each circuit $C$ in $\Lambda$ the number of
triangular faces of $\Lambda^\prime$ that are interior to $C$ is independent
of the triangulation $\Lambda^\prime$.}

{\it Proof:}  The result follows from (2.2) since $\ell$ and $v$ do not
depend on the chosen triangulation $\Lambda^\prime$. \lanbox

With the aid of triangulation we can describe the {\bf canonical flux
distribution} for {\it any\/} planar graph $\Lambda$.  Choose any
triangulation $\Lambda^\prime$ and place flux $\pi /2$ in every
triangular face.  By Lemma 2.2, this defines phases $\theta (x,y)$ on
$E(\Lambda^\prime)$ and hence on $E(\Lambda)$.
{\it A-priori}, these phases might depend on
the triangulation but, by Corollary 2.4, all triangulation give rise to
the same set of fluxes through the circuits of $\Lambda$.  By Lemma 2.1
the $\theta (x,y)$'s are uniquely defined up to a gauge transformation,
i.e., $\theta (x,y) \rightarrow \theta (x,y) + \phi_x - \phi_y$ with
the function $\phi_x$ being the only quantity that might depend on the
triangulation. Since the flux distribution is invariant under gauge
transformations, the canonical flux distribution is well defined!

Of special interest to us are bipartite planar graphs.  In general a
{\bf bipartite graph} is a graph $\Lambda$ whose vertex set $V$ is
the union of two disjoint sets $A$ and $B$ with the property that
$(x,y)$ is never an edge of
$\Lambda$ if $x \in A$ and $y \in A$ or $x \in B$ and $y\in B$.  We shall
assume $\vert B \vert \geq \vert A \vert$.  If
$\Lambda$ is a planar bipartite graph the canonical flux will always be
$\pi$ through every elementary square, zero through every elementary
hexagon etc.  However one has to be cautious about this because one could
have, for instance, a square with vertices $a,b,c,d$ and a fifth vertex $g$
inside the square connected by an edge only to $a$.  In this case our rule
says that the canonical flux through the circuit $a,b,c,d$ is zero and not
$\pi$.

A special feature of bipartite graphs, planar or otherwise, is that the
nonzero eigenvalues of any hopping matrix $T$ come in opposite pairs, i.e.,
if $\lambda$ is an eigenvalue of $T$
then so is $-\lambda$.  This follows from
$T = - U^* TU$ where $U$ is the diagonal unitary matrix with $+1$ on the
$A$-sites and $-1$ on the $B$-vertices.  $T$ itself can be written in the form
$\pmatrix{0&M\cr M^*&0\cr}$, where $M$ contains the matrix elements between
$A$ and $B$ sites.
\bigskip\noindent
{\bf III.  DETERMINANTS FOR PLANAR GRAPHS}

One of the main theorems of this paper is Theorem 3.1 about
determinants of bipartite graphs, and one of the concepts needed there
is that of the {\bf dimer partition function} $D(T)$ of the
graph $\Lambda$ with hopping matrix $T$.  A {\bf dimer covering} (or
matching) of $\Lambda$ is a subset $\{ e_1, e_2, \dots, e_n \}$ of $E$
such that every site in $\Lambda$ is an end point of precisely one of
the $e_i$'s.  In general, $\Lambda$ has many dimer coverings, but it
may have none at all.  In particular, if $\vert \Lambda \vert$ is odd
or if $\Lambda$ is bipartite and $\vert A \vert \not= \vert B \vert$
then there are no dimer coverings.

We define the {\bf dimer partition function} to be
$$D(T) = \sum
\limits_{{\rm dimer \ coverings}} \prod \limits_i \vert t_{x_i, y_i}
\vert, \eqno(3.1)$$
where the product is over all the edges $e_i =
(x_i, y_i)$ that constitute a particular dimer covering.  If $\vert
t_{xy} \vert=1$ then $D(T)$ is just the number of dimer coverings of
$\Lambda$.  Note that $D(T)$ depends only on the $\vert t_{xy} \vert$'s
and is therefore independent of the fluxes. In particular, $D(T)$ is
determined by the upper triangular array $\{t_{xy}\}_{x\le y}$. (See the
appendix.)

{\bf 3.1.  THEOREM (Canonical flux  counts dimers and maximizes
bipartite graph determinants).}  {\it Let $\Lambda$ be a planar graph
and let $\vert t_{xy}\vert$ be given positive numbers for all edges
$(x,y)$ in $\Lambda$.  For the canonical flux distribution
$$\det T =
(-1)^{\vert \Lambda \vert /2} D(T)^2. \eqno(3.2)$$
If, in addition,
$\Lambda$ is bipartite the canonical flux distribution maximizes $\vert
\det T \vert$ among all flux distributions.}

Before proving the theorem we make a series of remarks:

(i).  Unless $\vert A \vert = \vert B \vert$ in the
bipartite case, $D(T)=0$ and $\det T = 0$ for {\it every\/} choice of
flux.  In the general case, $D(T)=0$ unless $\vert \Lambda \vert$ is even.
More generally, we could consider non-bipartite graphs with $T$ of the form
$T_K = \pmatrix{0&M\cr M^* &K\cr}$, with $K$ selfadjoint.  This means that
edges are added between $B$-vertices but not between $A$-vertices.  It is
then an easy exercise in linear algebra to prove that $\det T_K = 0$ unless
$\vert B \vert \geq \vert A \vert$, and that if $\vert B \vert = \vert A
\vert$ then $\det T_K$ is independent of $K$, i.e., $\det T_K = \det T_0$.
As an example, start with the simple square, i.e., $\vert \Lambda
\vert = 4$ and (1,2), (2,3), (3,4), (4,1) are the edges.  Theorem 3.1 says
that the determinant is maximized by flux $= \pi$ through the square.  Now
add a diagonal edge (1,3) with some hopping amplitude $\vert t_{1,3} \vert$
on this new edge.  We now have a graph that consists of two triangles.  The
observation just made says that the determinant is independent of the
individual fluxes through the two triangles and depends only on their sum.
The canonical flux distribution, which is $\pi /2$ in each triangle, is
optimal, but is by no means the unique optimizer.

(ii).  In the bipartite case, the sign of the determinant as given in
(3.2) is correct for any $T$, not just the canonical $T$. This follows
from the $\lambda, -\,\lambda$ pairing of the eigenvalues which holds for
a bipartite lattice.

We can go further and define the elementary symmetric functions
$$e_k (T) = \sum \limits_{i_1 < i_2 < \dots < i_k} \prod \limits^k_{j=1}
\lambda_{i_j} (T) \eqno(3.3)$$
for $1 \leq k \leq \vert \Lambda \vert$ and with $e_0 (T) = 1$.  They are the
coefficients in the characteristic polynomial
$$\det (T + z) = \sum
\limits^{\vert \Lambda \vert}_{k=0} e_k (T) z^{\vert \Lambda \vert
-k}. \eqno(3.4)$$
Here, the
$\lambda_i (T)$'s are the eigenvalues of $T$.  For the same reason (the
$\lambda, - \lambda$ pairing) we see that for a bipartite graph
$$\eqalignii{e_k (T) &= 0, \qquad &k \ {\rm odd} \cr
e_k (T) &= (-1)^{k/2} \vert e_k (T) \vert, \qquad &k \ {\rm even}.
\cr}\eqno(3.5)$$
In fact, $e_k (T)$ is a sum of determinants of principal submatrices of $T$
in which $\vert \Lambda \vert - k$ columns and corresponding rows are
removed.  Each such submatrix is naturally associated with a subgraph of
$\Lambda$ --- which is necessarily bipartite as well.  Therefore, we can
conclude that the sign of the determinant of every principal submatrix of
even order, $k$, is $(-1)^{k/2}$.  For odd $k$, such determinants are
always zero.  Warning:  The canonical flux distribution need not maximize
$\vert e_k (T) \vert$ for $k \not= \vert \Lambda \vert$.  The reason is
that the canonical flux distribution for a subgraph might differ from the
one for the full graph $\Lambda$.  See Theorem 5.1, however.

(iii). The canonical flux distribution maximizes $\det T$ in the
bipartite case. It fails, generally, to do so in the non-bipartite case;
nevertheless, it does have a ``maximum property'' which is given in
Theorem A.2 in the appendix.

(iv).  In the Appendix it is shown that Theorem 3.1 is one of the two key
ingredients in a proof of Kasteleyn's theorem.

{\it Proof:}  By definition, the determinant is a sum over permutations of
monomials in the matrix elements of $T$, each of which is a product of the
kind $\varepsilon (\pi) t_{1,\pi (1)} \cdots t_{\vert \Lambda \vert, \pi
(\vert \Lambda
\vert)}$.  Here, $\varepsilon (\pi) =\pm 1$ is the signature of $\pi$.
Using the cycle decomposition of the permutation $\pi$ we see
that the above monomial can be written as $\prod \limits^k_{j=1}
(-1)^{\ell_j-1} \prod \limits_{C_j} T$, where $C_1, \dots , C_k$ is a family
of circuits with the property that every vertex of the graph is in precisely
one circuit.  Here $\ell_j$ denotes the length of the circuit $C_j$.  By the
definition of the canonical flux distribution, $\prod \limits_{C_j} T =
\prod \limits_{C_j} \vert t_{xy} \vert \exp [\pm i \pi f_j/2
]$, where $f_j$ is the number of interior triangles of $C_j$,
and the sign in the exponent indicates the orientation of $C_j$.
Thus, the determinant is now a sum over all circuit decompositions of terms
of the form $\prod \limits^k_{j=1} \prod \limits_{C_j} \vert t_{xy} \vert
(-1)^{\ell_j-1} \cos (\pi f_j/2)$.
Note that the factor 2 is counted by distinguishing circuits of different
orientations.

By Lemma 2.3, $f_j = \ell_j + 2v_j - 2$.  Thus, $(-1)^{\ell_j-1} \cos (\pi
f_j/2) = (-1)^{\ell_j-1} \cos (\pi \ell_j/2 + \pi v_j - \pi) =
(-1)^{\ell_j} \cos (\pi \ell_j/2 + \pi v_j)$.
If $\ell_j$ is odd, the cosine vanishes. Hence, {\it only even circuits
contribute to the determinant}. This is a crucial property of the
canonical flux distribution!  Moreover, since every vertex must belong
to a circuit, and every circuit has even length,  $v_j$ is also even for all
$j$ and hence $2v_j \equiv 0$ (mod 4) and does not contribute to the
sign of the monomial.  Therefore the monomial equals
$\prod \limits^k_{j=1}{\rm cos}(\pi \ell_j/2) \prod \limits_{C_j}
\vert t_{xy} \vert= \prod \limits^k_{j=1} (-1)^{\ell_j/2} \prod
\limits_{C_j} \vert t_{xy} \vert = (-1)^{\vert \Lambda \vert /2}
\prod \limits_{C_j} \vert t_{xy} \vert$ since
$\sum \nolimits^k_{j=1} \ell_j = \vert \Lambda \vert$. Note that when
$\vert \Lambda \vert$ is odd there is at least one circuit of odd length in
every circuit decomposition, and hence $\det T=0$.

The last step is to derive relation (3.2), which is geometrically
``obvious''.  It suffices to note in our case that $D(T) \cdot D(T)$ is
a sum of terms, each of which is of the form $\D_1 \D_2$, where $\D_1$
(likewise $\D_2$) denotes a single term in (3.1) corresponding to a
single dimer covering.  If we superimpose the two coverings we get a
collection of disjoint circuits $C_1, \dots, C_k$ on $\Lambda$.  Each site
of $\Lambda$ is in exactly one of these circuits.  Additionally, each circuit
will have an {\it even\/} length.  This ``circuit covering'' of $\Lambda$
corresponds to a term in $\det T$.  Conversely, each term in $\det T$
corresponds to a ``circuit covering''.  (Note: it is here that we use the
fact that only circuits of even length contribute to $\det T$, for
otherwise some terms in $\det T$ might give rise  to ``circuit coverings'' that
contain circuits of odd length.) All that is needed is to check that the
weights in $D(T)^2$ correspond to those in $\det T$.  The weight of a
``circuit covering'' in $\det T$ is $2^n$, where $n \leq k$ is the number
of circuits whose length exceeds 2.  The factor of 2 comes from the two
possible orientations of the circuit or, in other words, the contribution
of a cyclic permutation and its inverse.  The same factor $2^n$ arises
in $D(T)^2$ because each circuit can be decomposed into a dimer covering
of the circuit in exactly two ways.  \lanbox
\bigskip\noindent
{\bf IV. RINGS WITH ARBITRARY WEIGHTS}

We begin our study of the problem of maximizing eigenvalue sums of $T$
with respect to fluxes by considering the simplest
possible case.  In the process some notation and identities will be
established that will prove useful in later sections of this paper.

A {\bf ring} of $R > 2$ vertices (or $R$ edges) is a graph $\Lambda$ with
$\vert \Lambda \vert = R$ vertices
labeled 1 up to $R$ and with edges $(1,2), (2,3), \dots , (R-1, R), (R,1)$.
The hopping matrix is then determined by $R$ complex numbers $t_{12}, \dots
, t_{R1}$ with magnitudes given a priori as $\vert t_{i,i+1} \vert$.  Note
that $\Lambda$ is not necessarily bipartite, i.e., $R = \vert \Lambda \vert$
does not have to be even.

Although the spectrum of $T$ is easy to compute explicitly if $\vert t_{i,
i+1} \vert$ is independent of $i$, and hence one might think that our main
theorem here, 4.1, is without content, we draw the reader's attention to the
fact that we shall consider all possible $T$'s.  In other words, we shall
be dealing with the ``random one-dimensional Laplacian'' whose spectrum is
the object of much current research.  From this point of view, it is
somewhat surprising that some physical quantities of this random system can
easily be maximized with respect to the flux.

While our goal is to compute $E_0 (T)$ in (1.2),
we shall consider more
general functions of the eigenvalues of $T$.  Let $f: \R^+ \rightarrow \R$
be a real valued function defined for nonnegative reals, and define $F$ by
$$F(T) = \Tr f(T^2) = \sum \limits^{\vert \Lambda \vert}_{j=1} f
(\lambda^2_j) ,\eqno(4.1)$$
where $\lambda_1 \geq \lambda_2 \geq \dots \geq \lambda_{\vert \Lambda
\vert}$ are the eigenvalues of $T$.  The $f$ needed for
$E_0$ is
$$f_1(x) = \sqrt x \eqno(4.2)$$
while for $\ln \vert \det T \vert$ it is
$$f_2(x) = \mfr1/2 \ln x. \eqno(4.3)$$
Still another physically important function is
$$f_3(x) = \ln \cosh \sqrt x. \eqno(4.4)$$
appropriate to the free energy, $\F = - \beta^{-1} \ln \Xi$, in the grand
canonical
ensemble\footnote{$^\dagger$} {\baselineskip=12pt\sevenpoint In (4.5) we
have set the chemical potential $\mu$ equal to zero.  For a bipartite lattice
this yields an average particle number $M = 2N = \vert \Lambda \vert$, which
follows from $M = 2\sum\nolimits^{\vert \Lambda \vert}_{j=1} \exp (- \beta
\lambda_j) [1 + \exp (-\beta \lambda_j)]^{-1}$ together with the $(\lambda,
- \lambda)$ pairing.} (1.3):
$$\eqalignno{\F &= - 2\beta^{-1} \Tr \ln (e^{-\beta T} +1) = - 2\beta^{-1} \Tr
\{ \ln [\cosh (\beta T/2)] + \ln 2 - \beta T/2\} \cr
&= - 2\beta^{-1} \Tr f_3 (\beta^2 T^2/4) - 2\beta^{-1} \vert \Lambda \vert \ln
2
\qquad&(4.5)\cr}$$
since $\Tr\, T = 0$.  Here, $\beta^{-1} =$ (Boltzmann's constant) $\times$
(temperature).

All these functions have the property of being {\it concave}, i.e.,
$f(\lambda x + (1 - \lambda) y) \geq \lambda f(x) + (1 - \lambda) f(y)$ for
all $x \geq 0, y \geq 0$ and $0 < \lambda < 1$.  (In fact they are {\it
strictly\/} concave, i.e., equality implies that $x = y$.)

These three functions also belong to a more restricted class of functions
which we call {\bf integrated Pick functions}.  These are functions with the
integral representation
$$f(x) = c \ln x + \int \limits^\infty_0 \ln \left( 1 + {x \over s} \right)
\mu (\d s) \eqno(4.6)$$
where $c \geq 0$ and where $\mu$ is a nonnegative measure on $[0, \infty)$
such that this integral is finite.  For the functions in (4.2)-(4.4) we
have this integral representation with $\mu (\d s) = (\const.) s^{-1/2}$
d$s$ and $c=0$ for (4.2), $\mu (\d s) \equiv 0$ and $c = 1$ for (4.3) and
$\mu (\d s) =$\hfill\break
$\sum\nolimits^\infty_{k=0} \delta (s - [\pi (k + \mfr1/2)]^{-2}) \d s$,
and $c=0$ with $\delta =$ Dirac's $\delta$-function.  See [KL,
eqs. (3.12), (3.16)].  Combining (4.1) with (4.6) yields
$$F(T) = c \ln \det T^2 + \int \limits^\infty_0 \ln \det (1 + T^2/s) \mu
(\d s), \eqno(4.7)$$
and we see that the problem of maximizing $F(T)$ is reduced to that of
maximizing various determinants with respect to the flux.

The function $G(T)$ given in (1.4) cannot be represented in the form (4.7);
nevertheless we shall also be able to maximize $G(T)$.

{\bf 4.1.  THEOREM (Maximizing flux for the ring).}  {\it Consider a
hopping matrix $T$ with arbitrary, but fixed amplitudes $\vert t_{xy}
\vert$ on a ring of $R$ sites, let $f$ be an integrated Pick function
given by (4.6) and let $F(T)$ be as in (4.1).  Then the canonical flux
\hfill\break
$\pi
(R+2)/2\  (\mod 2 \pi)$ maximizes both $F(T)$ and the gap $G(T)$ if $R$ is
even.
If $R$ is odd, $F(T)$ and $G(T)$ are maximized by both of the choices $0$
and $\pi$.}

{\it Remark:}  When $R$ is odd the canonical flux is always $\pi/2$ or $3
\pi /2$ and never 0 or $\pi$.

{\it Proof:}  For $F(T)$ it suffices, by formula (4.7),
to show that the flux described
above maximizes $\det (c^2 + T^2) = \vert \det (ic + T) \vert^2$ for all
real numbers $c$.

As a first step we observe that for the invariants (or elementary symmetric
functions) we have that $e_k (T) = 0$ for $k$ odd and $1 \leq k
\leq R -1$.  This follows directly from remark (ii) after Theorem 3.1 when
$R$ is even.  When $R$ is odd it also follows from remark (ii) together
with the observation that every proper subgraph of a ring is bipartite.  We
also see, from remark (ii), that the sign of $e_{2m} (T)$ is $(-1)^m$.

With this information about the signs of the $e_k$'s, we can write, from
(3.4) with $z = ic$,
$$\det (c^2 + T^2) = \cases{\left( \sum \limits^{\vert \Lambda \vert/2}_{m=0}
\vert e_{2m}
(T) \vert c^{\vert \Lambda \vert -2m} \right)^2, &$\vert \Lambda \vert$
even \cr
\left( \sum \limits^{(\vert \Lambda \vert -1)/2}_{m=0} \vert e_{2m} (T)
\vert c^{\vert \Lambda \vert -2m}
\right)^2 + (\det T)^2, &$\vert \Lambda \vert$ odd .\cr} \eqno(4.8)$$
For future use we remark that both parts of (4.8),
hold for {\it any bipartite graph\/} $\Lambda$, not just a ring with
an even number of sites.

As a second step we shall show that in the expression (3.4) for $\det (T +
z)$, with $z \in \C$,
the invariants $e_k (T)$ are {\it independent of the flux\/} if $k < R$.
This is true only for a ring.  Note that the invariants are real
since $T$ is Hermitian.  Recall that the number $e_k (T)$ can be
computed from $T$ by
calculating the subdeterminants of $T$ with any $R-k$ columns and
corresponding rows removed, and then summing these numbers over all
possible removals.  The result is a sum of monomials of the form $\prod
\limits_j (-1)^{\ell_j-1} \prod \limits_{C_j} T$ where the product is
taken over all circuits $C_i$ that cover the subgraph obtained by removing
$k$ vertices and the corresponding edges.  But for $k < R$
the only circuits that
cover this subgraph form a dimer covering; their contribution does not
depend on the flux but only on the numbers $\vert t_{i, i+1} \vert$.  Thus
the only term in (3.4) that depends on the flux is $e_{\vert \Lambda \vert}
(T) = \det T$.

In both cases in (4.8), the problem of maximizing $\det (c^2 + T^2)$ is seen
to be the same as maximizing $\vert \det T \vert$.  If $R$ is even this
problem is solved in Theorem 3.1.  If $R$ is odd there are precisely two
circuits that contribute to $\det T$.  These are the circuits that traverse
the entire ring (in either direction) and correspond to an even
permutation.  Thus, for a ring of odd length
$$\det T = 2 Re \left\{ \prod \limits^R_{i=1} t_{i, i+1} \right\} = 2 (\cos
\Phi) \prod \limits^R_{i=1} \vert t_{i, i+1} \vert, \eqno(4.9)$$
from which we see that $\Phi =0$ or $\Phi = \pi$ maximizes $\vert \det T
\vert$, and hence also $F(T)$.  This completes the proof for $F(T)$.

To compute $G(T)$ we return to (4.7) and write $Q_\Phi (\lambda) \equiv
\det (T - \lambda) = P (\lambda) + \det T$, with $P$ being a polynomial of
order $R$ whose coefficients are {\it independent\/} of the flux $\Phi$.
$P$ is even if $R$ is even and $P$ is odd if $R$ is odd.  We note that as
$\det T$ varies between its maximum and minimum values, $Q_\Phi$ always has
$R$ roots.  We leave it to the reader to verify the following with the aid
of a graph of $Q_\Phi (\lambda)$.  Even $R$:  The maximum separation
between $\lambda_{R/2}$ and $\lambda_{R/2+1}$ is achieved by making $\vert
Q_\Phi (0) \vert$ as large as possible.  Since $Q_\Phi (0) = \det T$, this
means
choosing $\Phi$ to make $\vert \det T \vert$ as large as possible --- as
stated in our theorem.  Odd $R$:  If $\det T = 0$, the eigenvalues of
$Q_\Phi$ are paired (because $P$ is odd).  Then $-\lambda_{(R-1)/2} =
\lambda_{(R+3)/2}$ and $\lambda_{(R+1)/2} = 0$.  Thus $G_+ := [
\lambda_{(R+1)/2} - \lambda_{(R+3)/2} ] = [\lambda_{(R-1)/2} -
\lambda_{(R+1)/2} ] = :G_-$ when $\det T = 0$.  As $\vert \det T \vert$
increases, either $G_+$ increases, $G_-$ decreases and $\lambda_{(R+1)/2} >
0$ or vice versa and $\lambda_{(R+1)/2} < 0$.  Thus $G = \max (G_+,
G_-)$ and this increases with $\vert \det T \vert$.  By (4.9) we see that
$\Phi = 0$ or $\Phi = \pi$ maximizes $\vert \det T \vert$. \lanbox
\bigskip\noindent
{\bf V.  TREES OF RINGS}

Most of the results in Theorem 4.1 for bipartite rings with arbitrary hopping
amplitudes $\vert t_{xy} \vert$ can be extended to a much larger class of
planar graphs.  Two special cases of this class are the ladders and the
necklaces; they are discussed in detail in the next section because even
stronger results can be obtained for them.  It was those two classes, in
fact, that were the origin of this work and that were reported in [LE].

A planar graph $\Lambda$ is said to be a {\bf tree of rings} if and only if
$\Lambda$ has an embedding in $\R^2$ such that every circuit in $\Lambda$
has no interior vertices.

The simplest example consists of two rings which have exactly one vertex in
common.  Another example consists of two rings that have exactly one edge
(i.e., two neighboring vertices) in common.  More generally, one can have a
``tree of rings'' in which two successive rings share either one edge or
one vertex.  The canonical flux distribution for a tree of rings would have
flux $(\pi /2) [(\ell - 2) (\mod 4)]$ in each circuit of length $\ell$.

{\bf 5.1.  THEOREM (Maximizing flux for bipartite trees of rings).}  {\it
Let $\Lambda$ be a bipartite, planar graph that is a tree of rings and let
$\vert t_{xy} \vert$ be arbitrary given hopping amplitudes.  For $f$ an
integrated Pick function, let $F(T)$ be as in (4.1).  Then the canonical
flux distribution maximizes $F(T)$.  Moreover, it also maximizes the
magnitude of each elementary symmetric function $e_k (T)$ defined in
(3.3).}

{\it Proof:}  As in the proof of Theorem 4.1 ({\it cf}. eq. (4.8)) we have
that $\det (c^2 + T^2)$ will be maximized if we can {\it simultaneously}
maximize all the $\vert e_k (T) \vert$'s and if they all have the sign
$(-1)^{k/2}$.  The latter question was dealt with in the remark (ii) just
after Theorem 3.1.

Each $e_k (T)$ can be evaluated as a sum of determinants of principal
submatrices of $T$ of order $k$.  In terms of graphs, a particular term in
the sum is the determinant of $T$ restricted to a subgraph $\Lambda^\prime$
with $\vert \Lambda^\prime \vert = k$.  The important point is that the
circuits of $\Lambda^\prime$ are (i) a subset of the circuits of $\Lambda$
and (ii$^\prime$) they have no interior points.  The canonical flux
distribution for $\Lambda^\prime$ is the same as for $\Lambda$; this means
that if $C$ is a circuit that is both in $\Lambda$ and in $\Lambda^\prime$
then $\Phi_C = \Phi^\prime_C = 0$ or $\pi$ where $\Phi_C$ is the canonical
flux through $C$ (in $\Lambda$) and $\Phi^\prime_C$ is the canonical flux
(in $\Lambda^\prime$).  (Note:  The only way in which $\Phi_C$ could differ
from $\Phi^\prime_C$ is if $C$ had some interior vertices that were removed
in passing from $\Lambda$ to $\Lambda^\prime$.  But $C$ had no interior
vertices to start with.)  Hence each subdeterminant appearing in $e_k (T)$
is maximized (in absolute value) by the original canonical flux
distribution in $\Lambda$.  Since the signs of all these subdeterminants
are the same, in fact they depend only on $k$ (see remark (ii) after
Theorem 3.1), we see that $\vert e_k (T) \vert$ is maximized. \lanbox
\bigskip\noindent
{\bf VI.  LADDERS AND NECKLACES}

Most, but not all the graphs considered in this section are special cases
of those discussed in Section 5.  Here we consider certain graphs that are
finite subsets of the infinite lattice $\Z^2$, which is the infinite
embedded graph whose vertices are points in the plane with integer
coordinates and whose edges are the horizontal and vertical line segments
joining vertices a unit distance apart.  Of particular importance are {\bf
boxes}, which are the subgraphs of $\Z^2$ with 4 vertices and 4 edges
forming a circuit.  In general, our graphs need not be subgraphs of $\Z^2$,
i.e., they need not contain {\it all\/} the edges of $\Z^2$ that connect
the vertices $V$ in our graph.  (Example:  $\Lambda$ contains the 4
vertices of a box but only 3 of its edges.)  Evidently, all our graphs are
bipartite --- with the $A$ --- $B$ decomposition of their vertices being
the one
inherited from $\Z^2$.  Before describing them in detail, a few remarks are
needed.

If $T$ is a hopping matrix of a {\it bipartite\/} graph $\Lambda$, the
matrix $T^2$ is evidently block diagonal, i.e. $T^2 = \pmatrix{\alpha_T
&0\cr 0&\beta_T\cr}$, where $\alpha_T$ is $\vert A \vert \times \vert A
\vert, \ \beta_T$ is $\vert B \vert \times \vert B \vert$ and both are positive
semidefinite.  Assuming that $\nu := \vert B \vert - \vert A \vert \geq
0$, we have that the eigenvalues satisfy
$$\spec (\beta_T) = \spec (\alpha_T) \cup \{ \nu \ {\rm zeros} \}.
\eqno(6.1)$$
This is a simple consequence of the fact that $T = \pmatrix{0 &M\cr M^*
&0\cr}$, so that $\alpha_T = M M^*$ and $\beta_T = M^*M$.  Since the
eigenvalues of $T$ come in pairs, and there are $\vert \Lambda \vert$ of
them, we conclude that
$$\eqalignno{\spec (T^2) &= \{ \spec (\alpha_T) \ \hbox{with double
multiplicity}\} \cup \{\nu \ \hbox{zeros}\}, \qquad&(6.2)\cr
\spec (T) &= \spec \left( \sqrt{\alpha_T} \right) \cup \spec \left( -
\sqrt{\alpha_T} \right) \cup \{ \nu \ \hbox{zeros} \}. \qquad&(6.3)\cr}$$
Thus $\spec (T)$ is determined by {\it either\/} $\alpha_T$ or $\beta_T$
alone.

The matrix $\alpha_T$ has diagonal elements, $(\alpha_T)_{aa} = \sum
\nolimits_{b \in B}
\vert t_{ab} \vert^2$ for $a \in A$.  The off-diagonal elements of
$\alpha_T$
can be thought of as a hopping matrix of a new graph $\Lambda_A$, which
need not be planar.  The vertices of $\Lambda_A$ are the $A$-vertices of
$\Lambda$.  A necessary condition for the pair $(a, a^\prime)$ to be an
edge of $\Lambda_A$ is that there is a $B$-vertex, $b$, such that $(a,b)$
and $(b, a^\prime)$ are edges in $\Lambda$.  There may be more than one
such $b$ for a given pair $(a, a^\prime)$ and it is important to note that
since $(\alpha_T)_{aa^\prime} = \sum\nolimits_b t_{ab} t_{ba^\prime}$, it can
happen that $(\alpha_T)_{aa^\prime} = 0$, in which case $(a,a^\prime)$ is {\it
not\/} an edge of $\Lambda_A$, in conformity with our earlier convention.
Similar remarks hold for $(\beta_T)_{bb^\prime}$ and $\Lambda_B$.

There are special edges in $\Lambda_A$ or $\Lambda_B$ which we call {\bf
interior diagonals}.  These are edges $(a,a^\prime)$ or $(b,b^\prime)$ in
which $a$ and $a^\prime$ (or $b$ and $b^\prime$) belong to some (same) box
$S$ that is a subgraph of $\Lambda$.  This set of edges is denoted by
$D_A$.  Since circuits of length 4 can only be the edges of boxes, it
follows that the interior diagonals are the {\it only\/} edges that can
possibly disappear from $\Lambda_A$ because of the equality $\sum
\nolimits_b t_{ab} t_{ba^\prime} = 0$.  Likewise for $\Lambda_B$ and $D_B$.

The graphs we shall consider here can be described as follows.  Let
$\Lambda^\prime_A = \Lambda_A \sim D_A$, i.e., the vertices of
$\Lambda^\prime_A$ are those of $\Lambda_A$ but the edges are those of
$\Lambda_A$ {\it without\/} the interior diagonals.  Analogously,
$\Lambda^\prime_B$ is defined.  We say that $\Lambda$ is a {\bf hidden
tree} if either $\Lambda^\prime_A$ or $\Lambda^\prime_B$ is a tree (i.e.,
does not contain any circuits).

Two important examples introduced in [LE] are {\bf ladders} and {\bf
necklaces}.  Each is a connected union of $n$ boxes, labelled $1,2, \dots
, n$, forming a one-dimensional array.  In the ladder the boxes are joined
along (parallel) edges, with box $j$ connected to $j+1$.  In the necklace,
boxes $j$ and $j+1$ have only a single vertex in common, and these vertices
are either all $A$ or all $B$.  In both examples, boxes $j$ and $k$ are
disjoint if $\vert j-k \vert > 1$.  With the usual orientation of $\Z^2$,
ladders are either horizontal or vertical, while a necklace runs at
45$^\circ$ to either of these directions.

One can generalize the ladder by allowing occasional 90$^\circ$ bends,
while still keeping the one-dimensional character.  Now, squares $j$ and
$k$ can now have a vertex (but not an edge) in common if $\vert j-k \vert
= 2$; squares 1 and
$n$ are disjoint.  The bends cannot be completely arbitrary, however,
because the ``hidden tree'' condition must be maintained.  It can never be
maintained for a necklace with 90$^\circ$ bends.

Another, not so trivial example is that in which $\Lambda$ is the union of
4 squares, all of which have a vertex in common and which together form a
square of side length 2.  Here $\vert A \vert = 4$ and $\vert B \vert = 5$.
Although $\Lambda^\prime_A$ is a tree, $\Lambda^\prime_B$ is {\it not\/} a
tree.  (6.1) notwithstanding, it is somewhat surprising, when viewing the
graphs for $\Lambda_A$ and $\Lambda_B$, that they have the same spectrum
--- except for one zero eigenvalue.  Another example in
this vein is the $\Lambda$
that resembles a (3,2) Young diagram, i.e. 3 squares in a horizontal row
and 2 squares, also in a horizontal row, directly beneath them.  The
simplest case that is not a hidden tree, and for which none of our theorems
apply, is two rows of three squares each.

Finally, we make some remarks about the next theorem.

(i).  Ladders and necklaces are trees of rings, but the last three examples
(two rows of squares) are not.  Thus, for ladders and necklaces, the fact
that the canonical flux distribution maximizes $\Tr \vert T \vert$, $\F$ and
$\vert \det T \vert$ is already covered by Theorem 5.1.  The statement
about the gap is new, however, as is the method of proof.

(ii).  The concave function $F(T^2)$ is definitely a generalization of the
function in (4.1).  Not all concave functions (even those that are
invariant under unitary transformations) are eigenvalue sums as in (4.1).
In particular, the sum of the $k$ lowest eigenvalues of $T^2$ is not such a
function and it is needed, in fact, to prove the theorem about the gap.

{\bf 6.1.  THEOREM (Canonical flux maximizes concave functions on
hidden trees).}  {\it Let $\Lambda$ be a graph that is a subset of $\Z^2$
and suppose that $\Lambda^\prime_A$ (resp. $\Lambda^\prime_B$) is a tree.
Let $F$ be a concave function on the cone of positive definite matrices of
order $\vert A \vert$ (resp. $\vert B \vert$) with the property that
$F(U^*PU) = F(P)$ for every $P > 0$ and every gauge transformation $U$
(restricted to $\Lambda_A$, of course).  Finally, let $\{ \vert t_{xy}
\vert \}$ be unit hopping amplitudes on $\Lambda$, i.e. $\vert t_{xy} \vert
= 1$ if $(x,y) \in E(\Lambda)$.

Our conclusion is that among all hopping matrices $T$ with this unit
hopping amplitude, $F(\alpha_T)$ (resp. $F(\beta_T)$) is maximized by
the canonical flux.}

{\it Proof:}  We shall assume $\Lambda^\prime_A$ is a tree.  The proof for
$\Lambda^\prime_B$ is similar.  Assume $T$ maximizes $F(\alpha_T)$.  Let $U
= \{ u_x \delta_{xy}\}_{x,y \in V(\Lambda)}$ be the following gauge
transformation:  If $x = (n,m)$ with $n,m \in \Z$, then $u_x = (-1)^n$.
Let $Y = U^* TU$.
By concavity and gauge invariance, the block diagonal matrix $P^2 := \mfr1/2
(T^2 + Y^2)$ satisfies
$F(\alpha_P) \geq \mfr1/2 F(\alpha_T) + \mfr1/2 F(\alpha_Y) = F(\alpha_T)$.
This inequality proves our theorem if we can show that the matrix $\alpha_P$
can be achieved by the canonical flux distribution, i.e. if there is a gauge
such that $C:=$ ($T$ with the
canonical flux distribution) satisfies $\alpha_C = \alpha_P$.

Now note that $T^2$ and $Y^2 = U^*T^2U$ are related as follows:
$$\eqalignii{(Y^2)_{xy} &= - (T^2)_{xy} \qquad &\hbox{if} \ (x,y) \
\hbox{is an interior diagonal} \cr
(Y^2)_{xy} &= (T^2)_{xy} \qquad &\hbox{otherwise}. \cr}$$
Therefore, $(P^2)_{xy} = 0$ for $(x,y) \in D_A \cup D_B$ and $(P^2)_{xy} =
(T^2)_{xy}$ otherwise.

For any gauge, $(C^2)_{xy} = 0$ if $(x,y)$ is an interior
diagonal.  This is so because the flux through each box in $\Lambda$ is $\pi$,
and if we label the four vertices 1, 2, 3, 4 (counterclockwise) with 1 and
3 $\in V_A$ we have $(C^2)_{13} = C_{12} C_{23} + C_{14} C_{43}$.  But
$C_{12} C_{23} C_{34} C_{41} = -1$ and $C_{14} = \overline C_{41} =
1/C_{41}, C_{43} = 1/C_{34}$ (since, e.g., $\vert C_{14} \vert^2 = 1$), so
$(C^2)_{13} = 0$, as required.  As for the diagonal elements, they are
clearly equal, i.e., $(C^2)_{xx} = (P^2)_{xx}$.

Finally we have to compare the other matrix elements of $C^2$ and $P^2$ on
$\Lambda_A$.  In fact, they are both nonzero only on $\Lambda^\prime_A$, in
which case they satisfy $\vert (C^2)_{aa^\prime} \vert = \vert
(P^2)_{aa^\prime} \vert = 1$ because there is precisely one path (i.e.,
$B$-vertex) between $a$ and $a^\prime$.  Therefore $(C^2)_{aa^\prime} = \exp [i
\theta (a,a^\prime)] (P^2)_{aa^\prime}$ for each edge in
$\Lambda_{A^\prime}$.  The relevant question is then the following:  Is
there a gauge transformation $U$ such that $(U^* C^2 U)_{aa^\prime} =
(P^2)_{aa^\prime}$?  In other words, can we find $u_x = \exp [i \phi_x
]$ such that $\phi_a - \phi_{a^\prime} = \theta (a,a^\prime)$ for
every edge $(a,a^\prime)$ in $\Lambda^\prime_A$?  Since
$\Lambda^\prime_A$ is a tree, the answer is trivially, yes.  All
one-forms on a tree are exact. \lanbox

{\it Applications:}  For a graph with hopping amplitudes satisfying the
hypotheses of Theorem 6.1 the canonical flux distribution yields:
\item{(a)}  The lowest ground state energy $E_0 (T)$ and free energy $\F (T)$
for all temperatures.
\item{(b)}  The largest $\vert \det T \vert$ and gap $\widetilde G(T)$.
\medskip\noindent
The functions $-E_0 (T), - \F (T)$ and $\log \vert \det T \vert$ are concave
since they are integrated Pick functions, which are concave in $T^2$ as
mentioned in Section IV.  The gap $\widetilde G(T)$ can be computed
from the matrices
$\alpha_T$ and $\beta_T$.  Assuming that $\nu \equiv \vert B \vert - \vert
A \vert \geq 0$ we see that $\widetilde G(T) = 2 (\inf \spec \
\alpha_T)^{1/2}$ and that
$\widetilde G(T) = 2 \left( \sum \limits^{\nu + 1}_{i=1}\gamma_i
\right)^{1/2}$,
where $\gamma_i$ denote the eigenvalues of $\beta_T$ arranged in
increasing order.  Of course we used that $\gamma_1 =\gamma_2 = \dots =
\gamma_\nu = 0$.  Now the sum of the first $k$ eigenvalues of a Hermitian
matrix $H$ is a concave function of $H$.  Moreover since $x \mapsto \sqrt
x$ is concave and increasing, we see that $\widetilde G(T)$ is a concave
function of
$\alpha_T$ (resp. $\beta_T$).  Moreover it is gauge invariant and hence
satisfies the assumptions of Theorem 6.1.  Note that we had to discuss the
above two formulas for $\widetilde G (T)$ since both possibilities
($\Lambda^\prime_A$ is a tree or $\Lambda^\prime_B$ is a tree) have to be
considered.

In the case of ladders and necklaces the result about $E_0 (T), \F (T)$ and
$\vert \det T \vert$ were covered by Theorem 5.1, but the examples with
two rows of boxes, cited above, were not covered.  (In fact, the two-rowed
examples {\it cannot\/} be extended to the full generality of Theorem 5.1; see
Section VIIB.)  For ladders and necklaces, the result about the gap is not
covered by Theorem 5.1.  In [LE] it was mistakenly
asserted in Theorem 1 that (a) and (b) hold for {\it fully\/} generalized
ladders and necklaces in which {\it arbitrary\/} 90$^\circ$ bends are
allowed.  Indeed, for $E_0 (T), \F (T)$ and $\vert \det T \vert$ this is
correct (by Theorem 5.1).  For the gap, however, we must use Theorem 6.1
and this fails for bent necklaces and for ladders with arbitrary bends.  It
does hold, however, for generalized ladders that are hidden trees.

{\it Additional remarks and examples:}  There are two more cases where the
concavity argument in the proof of Theorem 6.1 is applicable but where the
graph is not necessarily a hidden tree, or even planar.

{\it (i) Row of cubes:}  Instead of a row of squares as in the ladder, take
$\Lambda$ to be a row of cubes joined on their faces, i.e., neighboring
cubes have four edges and one face in common.  Such a graph is, in fact,
bipartite and planar, but it is not a hidden tree.  If there are $n$ cubes
then $\Lambda$ is the $4 \times n$ planar, square lattice with ``periodic
boundary conditions'' in one direction.  [Indeed, we can even make the row
of cubes into a torus (i.e., attach the first cube to the last), which is
the same thing as the $4 \times (n+1)$ planar, square lattice with periodic
boundary conditions in both directions; the following argument will
continue to work in this case provided $n$ is odd.]

We assume, as in Theorem 6.1, that $\vert t_{xy} \vert = 1$ for every edge.
The flux in every face can easily be arranged to be $\pi$ in the following
way.  Start with the face that cube 1 and cube 2 have in common and put
flux $\pi$ through it by making $t_{xy} = 1$ on three edges and $-1$ on the
fourth.  Then use the negative of this on the corresponding
edges that cube 2 and cube 3
have in common --- and so on alternately.  Finally, set $t_{xy} = + 1$ on
the remaining edges, i.e., those edges that are perpendicular to the faces
between the cubes.  A first application of the concavity argument shows
that we get an upper bound in terms of $T^2$, but without interior diagonals
on the faces common to all the cubes.  In a second, similar application of the
argument we can get an upper bound in terms of a matrix that has no interior
diagonals on any of the other faces of the cubes as well.  In fact the only
nonzero elements of $T^2$ that remain will consist of four independent
one-dimensional chains (or else, in the case of the torus, four independent
rings of length $n+1$ with zero flux through each ring).  Theorem 4.1 says
that if $n+1 \equiv 2 (\mod \ 4)$ the optimizing flux for such rings is zero
and hence the above concavity argument shows that our choice of fluxes
cannot be improved.  If $n + 1 \equiv 0 (\mod \ 4)$ the optimum choice for
such a ring is flux $\pi$.  This can be achieved by a slight modification
of our initial choice of the $t_{xy}$'s along the edges perpendicular to
the inter-cube faces.  Initially we chose them all to be $+1$ but now we
choose them all to be $+1$ {\it except\/} for the $n^{th}$ cube.  There we
choose $t_{xy} = - 1$ along the four perpendicular edges.

{\it (ii) $SU(2)$-valued fields:}  We have considered the case that $t_{xy}
= \vert t_{xy} \vert \exp [ i \theta (x,y)]$ with $e^{i \theta}$ the
unknown variable.  It is also amusing to replace $e^{i \theta}$, which is
in $U(1)$ by
a $2 \times 2$ matrix $U_{xy}$ in $SU(2)$.  In other words, $T$ becomes a
$2 \vert \Lambda \vert \times 2 \vert \Lambda \vert$ Hermitian matrix in
which each $t_{xy}$ (for $x,y \in V(\Lambda)$) equals a given number $\vert
t_{xy} \vert$ times an $(x,y)$-dependent element of $SU(2)$.  We require
$t_{xy} = t_{xy}^*$.  Theorem 6.1 goes through in this case when $\vert
t_{xy} \vert = 1$.  We do not know whether Theorem 3.1, for instance, can
be generalized to the $SU(2)$ case.

There is, however, an interesting special feature of $SU(2)$.  For the
ladder or row of cubes, we had to break the translation invariance from
period one to period two in order to achieve flux $\pi$ in each face, i.e.,
$T$ could not be made the same in every box but we had to translate by two
boxes (or cubes) in order to recover $T$.  With $SU(2)$ fields we can
achieve the optimal flux distribution with period one.  This means the
following.  We require that the product of the four $t_{xy}$'s around a
square face (which is now a matrix product, of course) is the matrix $-I \in
SU(2)$.  This can be achieved by placing $i \sigma^1$ along all horizontal
edges, $i\sigma^2$ along all vertical edges and (in the case of cubes)
$i\sigma^3$ along all the edges in the remaining direction.  Here
$\sigma^1, \sigma^2, \sigma^3$ are the Pauli matrices.  We then have $-I$
in every face because, for example $(i\sigma^1) (i\sigma^2) (-i \sigma^1)
(-i \sigma^2) = -I$.
\bigskip\noindent
{\bf VII.  SOME CONJECTURES AND COUNTEREXAMPLES.}

{\bf (A).  The smallest determinant:}  A natural question, to be compared
with Theorem 3.1, is ``Which flux distribution {\it minimizes\/} $\vert \det T
\vert$ for a bipartite lattice?''  Since the canonical flux distribution
maximizes $\vert \det T \vert$ and since it places flux $\pi$ in each
square face (which is the maximum possible flux), it might be supposed that
the answer to the question is zero flux, i.e. set $t_{xy} = \vert t_{xy}
\vert$ for every edge in $E$.  In the case where $\Lambda$ is a simply
connected net of boxes on $\Z^2$ and $\vert t_{xy} \vert = 1$
the determinant was computed by Deift and
Tomei [DT] to have the three possible values $0, -1$ or $+1$.  Despite this
supportive example {\it the above conjecture is wrong\/}.  In the case of
two boxes with one common edge and $\vert t_{xy} \vert = 1$,
the determinant {\it vanishes\/} when the
flux in each square is $\pi /3$.  On the other hand, $\det T = -1$ when the
flux is zero.

{\bf (B)  The smallest energy:}  In Section V we have seen examples of some
graphs whose energy is minimized by the canonical flux distribution {\it
for arbitrary\/} hopping amplitudes.  Moreover, $\vert \det T \vert$ is
always maximized by that flux distribution for {\it every\/} bipartite,
planar graph with $\vert A \vert = \vert B \vert$.  For such an arbitrary
graph it is therefore natural to conjecture that $E(T)$ is also always
minimized by the canonical flux distribution.  Alas, this conjecture is
also false for arbitrary $\vert t_{xy} \vert$, as we how show by an
example.

In $\Z^2$ consider the graph consisting of four boxes arranged in a square,
i.e. $\Lambda$ has the nine vertices $o = (0,0), a = (1,0), b = (0,1), c =
(-1,0), d = (0, -1)$ and $(\pm 1, \pm 1)$.  If the conjecture were true for
arbitrary amplitudes then $E_0(T)$ would always be minimal if the flux in
each square is $\pi$.  If we now let $\vert t_{oa} \vert, \vert t_{ob}
\vert, \vert t_{oc} \vert$ and $\vert t_{od} \vert$ tend to zero, $E_0(T)$
becomes that of a ring of eight sites for which $E_0(T)$ is minimized by flux
$\pi$ (not $0 \equiv 4\pi$), as we saw in Theorem 4.1.  Indeed,
to see that 0 and $\pi$ do not give the same value for $E_0(T)$ in general,
for this ring, assume that the $t$'s on the ring all have amplitudes equal
to one.  If the flux is zero then $E_0(T) =- \Tr \vert T \vert$ is
easily computed to be $-2(1 + \sqrt 2)$ and for flux $\pi$ it equals
$-2^{5/4} (\sqrt{1 + \sqrt 2} + \sqrt{\sqrt 2 - 1})$ which is more
negative.
\bigskip\noindent
{\bf VIII.  THE FALICOV-KIMBALL MODEL}

So far all our models concerned hopping matrices on a graph $\Lambda$.  It
is likely that the results obtained do not hold if we add a diagonal term to
$T$, i.e., if we replace $T$ by the matrix
$$H = -T + 2UW. \eqno(8.1)$$
Here $W$ is a real diagonal matrix satisfying $0 \leq W_x \leq 1$ for each
$x$ and $U$ is a given real number called the coupling constant.

The eigenvalues of $H$ can be interpreted as  the possible energy levels of
a single electron hopping on a graph $\Lambda$ with kinetic energy $-T$ and
potential energy $2UW$.  This model, with $W_x$ restricted to be 0 or 1,
was introduced in [FK] as a model for a semiconductor-metal transition, and
it was studied extensively in [KL], where it was called the ``static
model'', and in [BS].  Our generalization to $W_x \in [0,1]$ for each $x$
is a mild one that we include here primarily because it can be handled
without extra complication.  The points of view taken in [KL] were
different from that in [FK].  There, the model was considered either as a
simplified version of the Hubbard model in which electrons of one sign of
spin are infinitely massive and therefore do not hop, or else as a model of
independent electrons interacting with static nuclei.  In the first view,
$U > 0$ and $U < 0$ are both relevant.  In the second, $U < 0$ is the
physically relevant sign, and $W_x = 1$ (resp. 0) denotes the presence
(resp. absence) of a nucleus at $x$.

The eigenvalues of $H$ in (8.1) are denoted by $-\nu$ (not $-\lambda$)
which,
as usual, are ordered $\nu_1 \geq \nu_2 \geq \dots \geq \nu_{\vert \Lambda
\vert}$.  The ground state energy of a system of $N$ {\it spinless\/}
electrons, interacting with a magnetic field and with the
nuclei, is given, as usual, by
$$E^{(N)}_0 = -\sum \limits^N_{i=1} \nu_i. \eqno(8.2)$$
(There is no 2 here, as in (1.1), because there is no spin.)
Notice that the spectrum of (8.1) is still invariant under gauge
transformations.  Thus the energy $E^{(N)}_0$ depends on $N$, the flux
distribution and, of course on $W$.

As mentioned before, minimizing the energy over all fluxes with a fixed $W$
is presumably a hopeless endeavor but the situation becomes easier if one
tries to minimize the energy with respect to the fluxes {\it and\/} $W$.
It is clear that the minimum is attained when $W = 0$ if $U > 0$ or when $W
= I$ if $U < 0$, which is uninteresting both mathematically and physically.
If, however, we introduce $N_n := \sum \limits_{x \in \Lambda} W_x =$
(total charge of the static particles), then the half-filled band condition
(from the Hubbard model point of view --- at least) is that $N_n + N =
\vert \Lambda \vert$.  This is the case that parallels the restriction in
the previous parts of this paper, since it means that on the average each
lattice site is occupied by one particle.   In fact we shall be a bit more
general in the $U > 0$ case, and will treat a slightly different case when
$U < 0$.  We shall optimize the energy over all flux distributions and all
potentials $W$ and all choices of $N$ subject to one of the following three
constraints.
$$\eqalignno{\sum \limits_{x \in \Lambda} W_x + N \leq 2 \vert A \vert
\qquad &\hbox{if} \ U < 0  \qquad&(8.3a) \cr
\sum \limits_{x \in \Lambda} W_x + N \leq
2 \vert B \vert \qquad &\hbox{if} \ U < 0 \qquad&(8.3b) \cr
\sum \limits_{x \in \Lambda} W_x + N \geq \phantom{2} \vert \Lambda \vert
\qquad
&\hbox{if} \ U > 0. \qquad&(8.3c) \cr}$$

Theorem 2.1 in [KL] says that in these three cases we can easily compute the
minimum of $E^{(N)}_0$ with
respect to $W$ and $M$ --- regardless of the flux distribution.  This result
requires only one particular structure of the graph, namely that it is
bipartite.  Nothing else is required.  As we shall see in Theorem 8.2 the
result is that the nuclei want to occupy only the $A$ sites or the $B$
sites in order to minimize the total energy.  We emphasie again
that this fact is {\it independent\/} of the magnetic field.

Lemma 8.1 and Theorem 8.2 are really a transcription to the $W_x \in [0,1]$
case of Lemma 2.2 and Theorem 2.1 in [KL].  Since they are short we give
them here.  It is convenient to introduce the matrix $S = 2W - I$, so $S$
is diagonal with $S_x \in [-1, 1]$.  Thus $H = h + UI$ with
$$h = -T + US .\eqno(8.4)$$
The matrix $h$ has eigenvalues $-\mu_1 \leq -\mu_2 \leq \dots \leq
-\mu_{\vert \Lambda \vert}$, with $\mu_j = \nu_j + U$.

{\bf 8.1.  LEMMA (Maximization with respect to the nuclear configuration).}
{\it Let $\Lambda$ be a bipartite (not necessarily planar) graph and $T$ a
prescribed hopping matrix on $\Lambda$.  We consider all functions $S$ on
the vertices of $\Lambda$ satisfying $-1 \leq S_x \leq 1$ for all $x$.  Let
$F$ be a concave, nondecreasing function from the set of Hermitian positive
semidefinite matrices into the reals.  Assume also that $F$ is gauge
invariant, $F(U^* PU) = F(P)$ when $U$ is a gauge transformation.

Then $F(h^2)$ is maximized with respect to $S$ at $S = V$ and at $S = -V$,
where
$$V_x = \cases{+1 &for $x \in A$ \cr
-1 &for $x \in B$ \cr}. \eqno(8.5)$$
If $F$ is strictly concave and strictly increasing then these are the only
maximizers.}

{\it Proof:}  The matrix $V = V^*$ is a gauge transformation and hence $h$
and $h^\prime := VhV$ satisfy $F(h^2) = F(h^{\prime 2})$.  Now $h^2 = T^2 +
U^2 S^2 - U(TS + ST)$ and, since $VTV = -T$ and $VSV = S$, $h^{\prime 2} =
T^2 + U^2 S^2 + U(TS + ST)$.  By concavity
$$F(h^2) = \mfr1/2 [F(h^2) + F(h^{\prime 2}] \leq F(T^2 + U^2 S^2) \leq F(T^2 +
U^2I), \eqno(8.6)$$
since $F$ is nondecreasing and $S^2 \leq I$.  Note that $T^2 + U^2 I =
h^2$ when $S$ is chosen to be $+V$ or $-V$.  If $F$ is strictly increasing and
strictly concave we can have equality in (8.6) only if $TS + ST = 0$ and
$S^2 = I$.  The former implies that $t_{xy} (S_x + S_y) = 0$, which implies
(since $\Lambda$ is connected) that $S =$ (constant)$V$.  The latter
implies that (constant) $= \pm 1$. \lanbox

{\bf 8.2.  THEOREM (Energy minima with respect to nuclear configurations).}
{\it Let $\Lambda$ be a bipartite graph (not necessarily planar) and let
$T$ be a prescribed hopping matrix.  We consider functions $W$ on the
vertices of $\Lambda$ satisfying $0 \leq W_x \leq 1$.

For the three cases given in (8.3) the minimum value of $E^{(N)}_0$
with respect to $W$ and $N$ is uniquely achieved as follows
$$\eqalignno{N = \vert A \vert \quad \hbox{and} \quad W = W_A := \mfr1/2 (I +
V), \phantom{N = \vert B \vert \quad \hbox{an}} \quad &U < 0 \qquad&(8.7a) \cr
N = \vert B \vert \quad \hbox{and} \quad W = W_B := \mfr1/2 (I - V),
\phantom{N = \vert B \vert \quad \hbox{an}} \quad &U < 0 \qquad&(8.7b) \cr
N = \vert B \vert \quad \hbox{and} \quad W = W_A \ \hbox{or} \ N = \vert A
\vert \ \hbox{and} \  W = W_B, \quad &U > 0. \qquad&(8.7c)\cr}$$

For each of these three cases, the minimum $E^{(N)}_0$ is given by
$$E^{(N)}_0 = - \mfr1/2 \Tr \vert h \vert + \mfr1/2 \Tr h + UN, \eqno(8.8)$$
where $h = -T + U (2 W - I)$.}

{\it Proof:}  By (8.2) and (8.4) we have that $E_0 = -\sum \nolimits^N_{j=1}
\nu_j \geq - \mfr1/2 \Tr \vert h \vert + \mfr1/2 \Tr h + UN = : A(h)$.  By
Lemma 8.1, $\Tr \vert h \vert = \Tr \sqrt{h^2}$ is maximal precisely at $W =
W_A$ or $W = W_B$ since $x \rightarrow \sqrt x$ is a strictly increasing
and strictly concave function.  Also, $\mfr1/2 \Tr h + UN = U \{ \sum
\nolimits_x W_x - \vert \Lambda \vert /2 + N \} =: B(h)$.  If $U > 0, B (h)
\geq \vert \Lambda \vert /2$.  If $U < 0 \ B(h) \geq U (2 \vert A \vert -
\vert \Lambda \vert /2)$ for (8.3a) and $B(h) \geq U (2 \vert B \vert -
\vert \Lambda \vert /2)$ for (8.3b).   These three lower bounds in $B(h)$
are attained (under conditions (8.3)) if $N$ and $W$ satisfy (8.7).

To complete the proof, we have to show that the lower bound on $- \mfr1/2
\Tr \vert h \vert = : C(h)$, given in Lemma 8.1, is compatible with the
condition on $N$ given in (8.7) (when $W$ is also that given in (8.7)).
For example, we have to show that if $W = W_A$ and $U < 0$ then the sum of
the negative eigenvalues of $h$ (namely $-\sum \nolimits_{\mu_j \geq 0}
\mu_j = - \mfr1/2 \Tr \vert h \vert + \mfr1/2 \Tr h)$ equals the sum of
the lowest $\vert A \vert$ eigenvalues of $h$.  In other words, we have to
show that $h$ has exactly $\vert A \vert$ negative eigenvalues.  We do so
now with a proof different from the one in [KL].  First note that for $t
\in [0,1], h_t = -tT + UV$ has no zero eigenvalues when $U \not= 0$ since
$h^2_t = t^2 T^2 + U^2 I \geq U^2 I > 0$.  Second, the matrix $h_0 = UV$
has precisely $\vert A \vert$ negative eigenvalues because $U < 0$.
Since the eigenvalues of $h_t$ are continuous functions of $t$ and because
no eigenvalue can cross zero, $h$ also has precisely $\vert A \vert$
negative eigenvalues.  The other two cases (8.7b) and (8.7c) are treated
in the same fashion.  \lanbox

The next theorem is our main result about the FK model in a magnetic field.

{\bf 8.3.  THEOREM (Canonical flux minimizes energy on trees of rings).}
{\it Let $\Lambda$ be a bipartite tree of rings, $T$ a hopping matrix with
arbitrarily prescribed amplitudes, $\vert t_{xy} \vert$.  As in Theorem 8.2
we consider functions $W$ on the vertices of $\Lambda$ satisfying  $0 \leq
W_x \leq 1$.

For the three cases given in (8.3) the minimum value of $E^{(N)}_0$ with
respect
to the flux distribution, $N$ and $W$ is achieved by the canonical flux
distribution together with the $N$ and $W$ given by (8.7).}

{\it Proof:}  Minimizing first with respect to $N$ and $W$, we can assume,
by Theorem 8.2, that (8.7) is satisfied.  In these cases we have that $E_0
= - \mfr1/2 \Tr \vert h_{A,B} \vert +$ constant, where the constant depends
on the case but {\it not\/} on the flux distribution.  In each case, $W =
W_A$ or $W_B$ and we denote the two choices of $h$ by $h_A$ and $h_B$.
Note that $N$ no longer enters the discussion.  Our only goal now is to
maximize $\Tr \vert h_{A,B} \vert$ with respect to the flux distribution.
But $\Tr \vert h_{A,B} \vert = \Tr \sqrt{h^2_{A,B}}$ and $h^2_{A,B} = T^2 +
U^2 I$ since $S^2_{A,B} = I$.  The function $x \mapsto \sqrt x$ is an
integrated Pick function, i.e., $\sqrt x = d \int \limits^\infty_0 \ln
(1 + x/s ) s^{-1/2} ds$ for some constant $d > 0$ (see.
4.6).  Hence maximizing $\Tr \vert h_{A,B} \vert$ is reduced to
maximizing $\det (c^2 + h^2_{A,B}) = \det (c^2 + U^2 + T^2)$ for all
constants $c$.  That this is achieved by the canonical flux distribution on
trees of rings is precisely the content of Theorem 5.1.  \lanbox
\bigskip\noindent
\vfill\eject
\bigskip\noindent
{\bf APPENDIX:  KASTELEYN'S THEOREM}

We give here a different, and we believe more transparent
proof of a deep theorem
due to Kasteleyn [KP], which is one of the main tools for counting dimer
configurations on planar graphs.  Let us emphasize that $\Lambda$ is now a
finite graph that is {\it not necessarily bipartite}.

Historically, the motivation behind Kastelyn's theorem was an attempt to
calculate efficiently the partition function $D(T)$ in (3.1)
for large planar
graphs--- by reducing the problem to the calculation of a determinant.
This was accomplished by Temperley and Fisher [TF] in special cases, but
independently and in full generality by Kasteleyn [KP]. The starting point
was Pfaff's theorem for an {\it antisymmetric matrix\/} $A$ (of even
order): $\det A ={\rm Pf}(A)^2$.  Here $\Pf(A)$ is the {\bf Pfaffian} of $A$
(more precisely, the Pfaffian of the upper triangular array of
$A=\{a_{xy}\}_{1\le x< y\le N}$), defined by
$${\rm Pf} (A) = \sum \limits_\pi \varepsilon (\pi) a_{\pi (1) ,\pi (2)}
a_{\pi (3) ,\pi (4)} \dots a_{\pi (N-1)} ,a_{\pi (N)}\eqno{(A.1)}$$
where the sum is over all permutations $\pi \in S_N$ with $\pi (1) < \pi
(3) < \pi (5) < \dots$ and with $\pi (i) < \pi (i+1)$ for odd $i$.  Also,
$\varepsilon (\pi)$ is the signature of $\pi$.

Each term in Pf$(A)$ corresponds to a dimer covering of the graph
$\Lambda$ with $N$ vertices and with edges corresponding to the nonzero
elements of $A$.

The Kasteleyn, Temperley-Fisher idea is to set $a_{xy} = \exp [i \theta
(x,y)] \vert t_{xy} \vert$ for $x < y$, with $\theta (x,y)$ chosen so that
all terms in (A.1) have a common argument $\theta$.  Then, trivially, $D(T)
= e^{i\theta} \Pf (A)$ for some $\theta$ and $\vert \det A \vert = D(T)^2$.

In Theorem 3.1 we solved the $D(T)$ problem, from a different perspective,
by using the canonical flux distribution:  $\vert \det T \vert = D(T)^2$.
Moreover, we gave a very simple rule (in the proof of Theorem 3.1 and in
the remark following Lemma 2.2) for an explicit construction of $\exp [i
(\theta (x,y)]$.  We did {\it not\/} try to construct an antisymmetric
matrix, but it is a fact that among the gauge equivalent $T$'s with
canonical flux distribution, there {\it is\/} one that is antisymmetric.
This is Theorem A.1 below.  We note here that $\det T$ is a gauge invariant
quantity, but $\Pf (T)$ is not.

Theorem A.1, together with Theorem 3.1 yields {\it Kasteleyn's theorem}, namely
that there is always a real, antisymmetric matrix $A$ such that $T = iA$
and $\Pf (A) = D(T)$.  To see this implication, note that Theorem (3.1)
says $\det A = \det (iT) = (-1)^{\vert \Lambda \vert /2} (-1)^{\vert
\Lambda \vert /2} D(T)^2$.  On the other hand $\det A = \Pf (A)^2$.  We can
then make $\Pf (A) = + D(T)$ by multiplying the first row and column of $A$
by $-1$, if necessary.  One way in which our proof is a little simpler than
Kasteleyn's is that graphs with cut-points do not require special
treatment, in either of Theorems A.1 or 3.1.

Theorem A.2 is another corollary of Theorem A.1.  It answers the question:
For which class of matrices (in the non-bipartite case) does the canonical
flux distribution maximize $\vert \det T \vert$?  Clearly the canonical
flux cannot maximize $\vert \det T \vert$ in general.  (A simple
counterexample is provided by the triangle, $\vert \Lambda \vert = 3$ where
$\vert \det T \vert$ is maximized by flux 0 or $\pi$ (Theorem 4.1)
but the canonical flux is $\pi /2$.)

{\bf A.1. THEOREM (Canonical flux and antisymmetry).}  {\it Let
$\Lambda$ be a finite planar graph with given hopping amplitudes $\vert
t_{xy} \vert$.
There exists a gauge such that the Hermitian hopping matrix $T$
defined by the canonical flux distribution has purely imaginary
elements, i.e., $T = iA$ with $A^T = -A$ and $A$ real. }

{\it Proof:}  We can assume $\Lambda$ is triangulated and we can start with
the remark following Lemma 2.2 which states (indeed, gives an explicit
rule) that there is a matrix $T^\circ$ whose fluxes are canonical and such
that $t^\circ_{xy} \in \{ i, -i, 1, -1 \}$ for all $(x,y) \in E$.  Edges
for which $t^\circ_{xy} = \pm i$ will be called ``good'' and those for
which $t_{xy} = \pm 1$ will be called ``bad''.  Our goal is to find a gauge
transformation $U_{xy} = u_x \delta_{xy}$ with $u_x \in \{ i, 1 \}$ for all
$x \in V$, such that $U^* TU$ has no bad edges.  We shall do so by showing
that if $T$ is {\it any\/} matrix with canonical fluxes and with $t_{xy}
\in \{ i, -i, 1, -1 \}$ and with at least one bad edge then we can find a
gauge transformation $U$ with $u_x \in \{ i, 1 \}$ such that $U^* TU$ has
at least one less bad edge than $T$ has.  Since the number of edges of
$\Lambda$ is finite, the theorem is proved by induction.

Let $(a,b) \in E$ be a bad edge.  Consider the set of sites $S = \{ y \in
V:$ there is a path from $a$ to $y$ whose edges are all good $\}$; by
definition $a \in S$.  Set $u_x = i$ if $x \in S$ and $u_x = 1$ if $x
\not\in S$.  Clearly, if $(c,d) \in E$ was a good edge for $T$ it remains a
good edge for $U^*TU$ (because either $u_c = u_d = 1$ or $u_c = u_d = i$).

To complete the proof we have only to show that the edge $(a,b)$ has become
a good edge.  Since $a \in S$, we have to show that $b \not\in S$.  Indeed,
suppose $b \in S$.  Then there is a circuit $C = a, x_1, x_2, \dots , x_n
b$, of length greater than 2, such that the edges $(a, x_1), (x_1, x_2)
\dots , (x_{n-1}, x_n)$ are good while $(x_n, b)$ is bad.  We claim that
this is impossible; in fact {\it every\/} circuit must have an
{\it even\/} number
of bad edges.  To see this, use eq. (2.2) in Lemma 2.3, which says that $f
= \ell \ (\mod 2)$.  Here, $f$ is the number of (triangular) faces inside
$C$.  We have $\Phi_C =$ flux through $C = \pm \pi f/2$ (by the definition
of the canonical flux distribution).  On the other hand, $\Phi_C = (\pi /2)
\sum \nolimits_G (\pm 1)$.  Here, the sum is over the good edges and $+1$
or $-1$ is taken according to the direction in which the edge is traversed
when $C$ is traversed in an anticlockwise sense.  In any event, $\Phi_C =
(\pi /2) \{ \vert G \vert (\mod 2) \}$ where $\vert G \vert$ is the number
of good edges in $C$.  Thus, $f = \vert G \vert (\mod 2)$, which proves our
assertion. \lanbox

{\bf A.2.  THEOREM (Canonical flux maximizes antisymmetric determinants).}
{\it Let $\Lambda$ be a planar (not necessarily bipartite) graph with
hopping amplitudes $\vert t_{xy} \vert$ given.  The canonical flux
distribution maximizes $\vert \det T \vert$ among all flux distributions
such that $T$ is both Hermitian and antisymmetric.}

{\it Proof:}  If $T$ is antisymmetric, $\det T = \Pf (T)^2$.  But, as we
see easily from the definition (3.1), $\vert \Pf (T) \vert \leq D(T)$.  By
Theorem 3.1, $\vert \Pf (T_c) \vert = D(T)$, where $T_c$ has the canonical
flux distribution.  By Theorem A.1, the gauge can be chosen so that $T_c$
is antisymmetric. \lanbox
\bigskip

{\baselineskip=3ex\eightpoint\smallskip
\font\eightit=cmti8
\font\eightbf=cmbx8
\def\it{\eightit}
\def\bf{\eightbf}

\bigskip\noindent
{\bf REFERENCES}
\item{[AM]}  I. Affleck and J.B. Marston, {\it Large n-limit of the
Heisenberg-Hubbard model:  Implications for high-$T_c$ superconductors},
Phys. Rev. {\bf B37}, 3774-3777 (1988).
\item{[BBR]}  A. Barelli, J. Bellissard and R. Rammal, {\it Spectrum of 2D
Bloch electrons in a periodic magnetic field:  algebraic approach}, J.
Phys. (France) {\bf 51}, 2167-2185 (1990).
\item{[BR]}  J. Bellissard and R. Rammal, {\it An algebraic semiclassical
approach to Bloch electrons in a magnetic field}, J. Phys. (France) {\bf
51}, 1803-1830 (1990).  J. Bellissard and R. Rammal, {\it Ground state of
the Fermi gas on 2D lattices with a magnetic field}, J. Phys. (France) {\bf
51}, 2153-2165 (1990).  J. Bellissard and R. Rammal, {\it Ground state of
the Fermi gas on 2D lattices with a magnetic field:  new exact results},
Europhys. Lett. (Switzerland) {\bf 13}, 205-210 (1990).
\item{[BS]}  U. Brandt and R. Schmidt, {\it Exact results for the
distribution of the f-level ground state occupation in the spinless
Falicov-Kimball model}, Z. Phys. {\bf B63}, 45-53 (1986).  U. Brandt and R.
Schmidt, {\it Ground state properties of a spinless Falicov-Kimball model;
Additional features}, Z. Phys. {\bf B67}, 43-51 (1987).
\item{[DT]}  P.A. Deift and C. Tomei, {\it On the determinant of the
adjacency matrix for a planar sublattice}, J. Comb. Theor. {\bf B35},
278-289 (1983).
\item{[FK]}  L.M. Falicov and J.C. Kimball, {\it Simple model for
semiconductor-metal transitions:  $SmB_0$ and transition metal oxides},
Phys. Rev. Lett. {\bf 22}, 997-999 (1969).
\item{[HD]}  D.R. Hofstadter, {\it Energy levels and wave functions of
Bloch electrons in a rational or irrational magnetic field}, Phys. Rev.
{\bf B14}, 2239-2249 (1976).
\item{[HLRW]}  Y. Hasegawa, P. Lederer, T.M. Rice and P.B. Wiegmann, {\it
Theory of electronic diamagnetism in two-dimensional lattices}, Phys. Rev.
Lett. {\bf 63}, 907-910 (1989).
\item{[HP]}  P.G. Harper, {\it Single band motion of conduction electrons
in a uniform magnetic field}, Proc. Phys. Soc. Lond. {\bf A68}, 874-878
(1955).  P.G. Harper, {\it The general motion of conduction electrons in a
uniform magnetic field, with application to the diamagnetism of metals},
Proc. Phys. Soc. Lond. {\bf 68A}, 879-892 (1955).
\item{[KG]}  G. Kotliar, {\it Resonating valence bonds and d-wave
superconductivity}, Phys. Rev. {\bf B37}, 3664-3666 (1988).
\item{[KP]}  P.W. Kasteleyn, {\it The statistics of dimers on a lattice I.
The number of dimer arrangements on a quadratic lattice}, Physica {\bf 27},
1209-1225 (1961).  P.W. Kasteleyn,
{\it Graph theory and crystal physics} in {\it Graph Theory and
Theoretical Physics}, F. Harary ed., Academic Press (1967), pp.
44-110.
\item{[KL]}  T. Kennedy and E.H. Lieb, {\it An itinerant electron model
with crystalline or magnetic long range order}, Physica {\bf 138A}, 320-358
(1986).
\item{[LE]}  E.H. Lieb, {\it The flux phase problem on planar lattices},
Helv. Phys. Acta {\bf 65}, 247-255 (1992).
\item{[RD]}  D.S. Rokhsar, {\it Quadratic quantum antiferromagnets in the
fermionic large-N limit}, Phys. Rev. {\bf B42}, 2526-2531 (1990).  D.S.
Rokhsar, {\it Solitons in Chiral-Spin liquids}, Phys. Rev. Lett. {\bf 65},
1506-1509 (1990).
\item{[TF]}  H.N.V. Temperley and M.E. Fisher, {\it Dimer problem in
statistical mechanics - An exact result}, Phil. Mag. {\bf 6}, 1061-1063
(1961).
\item{[WWZ]}  X.G. Wen, F. Wilczek and A. Zee, {\it Chiral spin states and
superconductivity}, Phys. Rev. {\bf B39}, 11413-11423 (1989).

}

\bye